\documentclass[twocolumn]{aastex7}

\usepackage{graphicx}
\usepackage{lineno}
\usepackage{multirow,tabularx}
\usepackage{array,ragged2e}
\newcolumntype{P}[1]{>{\RaggedRight\arraybackslash}p{#1}}
\newcolumntype{C}[1]{>{\RaggedCenter\arraybackslash}c{#1}}

\def\iso#1{$^{#1}$}

\begin{document}

\shorttitle{}
\shortauthors{Trueman et al.}

\title{Galactic chemical evolution of short-lived radioactive isotopes produced by explosive nucleosynthesis: $^{60}$Fe and $^{53}$Mn}

\correspondingauthor{Thomas Trueman}
\email{thomas.trueman@uni-bayreuth.de}

\author{Thomas C. L. Trueman}
\email{thomas.trueman@csfk.org}
\affiliation{University of Bayreuth, BGI, Universitätsstraße 30, 95447 Bayreuth, Germany}
\affiliation{Konkoly Observatory, HUN-REN Research Centre for Astronomy and Earth Sciences, H-1121 Budapest, Hungary}
\affiliation{CSFK, MTA Centre of Excellence, Budapest, Konkoly Thege Miklós út 15-17., H-1121, Hungary}

\author{Andr\'es Yag\"ue L\'opez}
\email{andyague@gmail.com}
\affiliation{Computer, Computational and Statistical Sciences (CCS) Division, Center for Theoretical Astrophysics, Los Alamos National Laboratory, Los Alamos, NM 87545, USA}
\affiliation{NuGrid Collaboration, \url{http://nugridstars.org}}

\author{Maria Lugaro}
\email{maria.lugaro@csfk.org}
\affiliation{Konkoly Observatory, HUN-REN Research Centre for Astronomy and Earth Sciences, H-1121 Budapest, Hungary}
\affiliation{CSFK, MTA Centre of Excellence, Budapest, Konkoly Thege Miklós út 15-17., H-1121, Hungary}
\affiliation{ELTE E\"{o}tv\"{o}s Lor\'and University, Institute of Physics and Astronomy, Budapest 1117, P\'azm\'any P\'eter s\'et\'any 1/A, Hungary}
\affiliation{School of Physics and Astronomy, Monash University, VIC 3800, Australia}

\author{Marco Pignatari}
\email{marco.pignatari@csfk.org}
\affiliation{Konkoly Observatory, HUN-REN Research Centre for Astronomy and Earth Sciences, H-1121 Budapest, Hungary}
\affiliation{CSFK, MTA Centre of Excellence, Budapest, Konkoly Thege Miklós út 15-17., H-1121, Hungary}
\affiliation{University of Bayreuth, BGI, Universitätsstraße 30, 95447 Bayreuth, Germany}
% \affiliation{E.A. Milne Centre for Astrophysics, Department of Physics \& Mathematics, University of Hull, HU6 7RX, UK}
\affiliation{NuGrid Collaboration, \url{http://nugridstars.org}}

\date{\today}

\begin{abstract}
Several short-lived radionuclides (SLRs) are know to have existed in the early Solar System (ESS). These species, which typically decay with half-lives of the order of a few million years, can be used to probe the timescale of events preceding the birth of the Sun. We investigate the ESS origin of $^{53}$Mn, produced by core-collapse (CCSNe) and Type Ia supernovae (SNe Ia), and $^{60}$Fe, produced exclusively by CCSNe. We model the evolution of the radioactive-to-stable abundance ratios of these SLRs with a galactic chemical evolution (GCE) framework accounting for different supernova yields, SN Ia delay times, and other galactic features $(K)$. A further set of models are calculated assuming that SN Ia did not contribute any $^{53}$Mn to the ESS. The predicted ratios are compared to meteoritic ratios to derive a distribution of solar isolation times that includes uncertainties due to stochastic chemical enrichment and precision of the ESS values. The isolation times are then compared to those of $^{107}$Pd and $^{182}$Hf calculated in previous work. A self-consistent solution can be found within the current uncertainties, especially when using the GCE setups with $K = 1.6$ and 2.3, although the maximum likelihood for the \iso{60}Fe distribution is typically $\sim 4-5$ Myr shorter than for \iso{53}Mn. The predicted \iso{60}Fe/\iso{53}Mn ratio, instead, is completely inconsistent with the ESS value; this could be resolved using a larger fraction of faint CCSNe than usually considered in GCE models.
  
\end{abstract}


\section{Introduction} 

Meteorites contain the chemical imprint of the Solar System during its nascent stages \citep{Kleine:2020}. Through their analysis, it is established that a number of radioactive isotopes with half-lives of $T_{1/2}\sim10^{5}-10^8$ yr existed in the early Solar System \citep[ESS; see e.g.][]{dauphas, lugaro18, Davis:2022}. These short-lived radionuclides (SLRs) can be used as \textit{cosmochronometers} for timescales relevant to the birth of the Sun, since the timescale of their decay is comparable to the time interval after which a Sun-like star may be expected to form inside a stellar nursery \citep{huss09, dauphas, lugaro18}, the so-called \textit{isolation time} (hereafter, $t_{\rm{iso}}$). Here, the Sun's birth time is defined by the formation of the first solids (the calcium-aluminium-rich inclusions, CAIs) in the early Solar System (ESS). 

In this work, we consider the ESS origin of two SLRs: $^{53}$Mn \citep[$T_{1/2}=3.74\pm0.04$ Myr;][]{Kondev:2021} primarily produced by Type Ia supernovae (SN Ia) with a secondary contribution from core-collapse supernovae (CCSNe); and $^{60}$Fe \citep[$T_{1/2}=2.62\pm0.04$ Myr;][]{Rugel:2009}, mostly produced by CCSNe \citep{limongi06}. Using galactic chemical evolution (GCE) models, we calculate the radioactive-to-stable abundance ratios for $^{53}$Mn/$^{55}$Mn and $^{60}$Fe/$^{56}$Fe in the interstellar medium (ISM) at the time of the birth of the Sun ($t_{\odot} \sim 8.5$ Gyr), exploring a wide range of initial input parameters. The GCE predicted abundance ratios are compared to the ESS ratios inferred from meteorites to help better understand the primitive environment of the Solar System. 

The abundances of manganese and iron in CAIs are low, therefore, the ESS abundances of \iso{53}Mn and \iso{60}Fe have been estimated by analysing later-forming chondrules and then time-correcting to the time of CAI formation \citep{Davis:2022}; the accuracy of this technique is limited by the uncertainty associated with the respective half-life $(T_{1/2})$ of a given SLR and the age of the samples. Initially, it was thought that $^{60}$Fe, together with \iso{26}Al, could have played an important role in the differentiation of planetesimals during the formation of the Solar System, due to the heat generated during its decay process \citep{Mostefaoui:2005}. However, revised estimates for the $^{60}$Fe/$^{56}$Fe ESS ratio are two orders of magnitude lower than previously thought \citep[the value of $1.01\pm0.27\times10^{-8}$ from][is adopted in this work]{Tang:2015}, thus essentially confirming $^{26}$Al as the only source of planetary heating in the ESS \citep{Tang:2012}. The $\gamma$-ray lines from $^{60}$Fe decay have been detected by INTEGRAL/SPI \citep{Harris:2005}, and are direct observational evidence for recent and ongoing nucleosynthesis; however, the observed emission is largely diffuse and lacks the spatial resolution needed to pinpoint individual star-forming regions \citep{wang:2020}. In contrast, no $\gamma$-ray is produced during the electron capture decay of $^{53}$Mn since most of the energy is imparted into the neutrino. The $^{53}$Mn/$^{55}$Mn ESS ratio used here was recalculated to be $7\pm1\times10^{-6}$ from high-precision U isotope measurements of primitive meteorite samples \citep{Tissot:2017}.  

It was first proposed by \cite{Cameron:1977} that a supernova could have triggered the collapse of the gas cloud that condensed to form the giant molecular cloud that nurtured the Solar System and injected the SLRs, specifically \iso{26}Al. Among the many alternative scenarios proposed since then \citep[see Figure 44 of][for a summary and references]{Diehl:2021}, the proto-solar cloud or the proto-planetary disk could have been later enriched via the explosive ejecta of one or several CCSNe from short-lived stars born within the same cloud as the Sun \citep[e.g.,][]{Hester:04, oue10}. However, a CCSN source would likely inject too much $^{60}$Fe into the presolar material \citep{vescovi2018,Battino:2024}, especially when considering the revised $^{60}$Fe/$^{56}$Fe ESS ratio obtained by, e.g., \cite{Tang:2012, Tang:2015} and \cite{Trappitsch:2018}. Furthermore, a CCSN would also lead to an overproduction of $^{53}$Mn, unless large amounts of the inner ejecta fallback onto the core \citep{takigawa}. A candidate for such a star could be a faint supernova, which experiences a large amount of fallback preventing the overproduction of $^{53}$Mn \citep{Meyer:2000, Lawson:2022}. Another currently favoured astrophysical candidate for a local source(s) injecting \iso{26}Al in the ESS are the winds of Wolf-Rayet stars \citep{Arnould:2006,Gounelle:2012} and/or massive binary stars \citep{Brinkman:2023}; however, these winds do not carry any \iso{55}Mn or $^{60}$Fe. Therefore, we explore the plausibility of the origin of the $^{53}$Mn and $^{60}$Fe in the ESS due to inheritance from the galactic background.
 
Early attempts to model the evolution of SLR abundances in the Milky Way used analytical models based on simplified mathematical frameworks \citep{clayton:1982, clayton84, Meyer:2000, wasserburg06,huss09}. These models often assumed a constant production ratio between an SLR and its stable reference isotope, and neglected the time-delayed contributions from different stellar sources. This approach is particularly inaccurate for $^{60}$Fe/$^{56}$Fe and $^{53}$Mn/$^{55}$Mn, as the isotopes involved have different stellar sources. More recently, numerical frameworks have been developed that include the stable and radiogenic yields from several different mass- and metallicity-dependent nucleosynthesis sources, as well as accounting for the time-delay between the formation of the progenitor and subsequent enrichment event \citep{Sahijpal:2014, Travaglio:2014, kaur:2019, cote19, Trueman:2022}. Using these, it is possible to calculate theoretical upper and lower limits for the SLR-to-stable abundance ratios in the ISM at $t_{\odot}$, taking into account different GCE input parameters, stellar modelling assumptions, and nuclear physics uncertainties. 

Due to the stochastic nature of stellar enrichment events and the fact that unstable isotopes undergo radioactive decay, the ISM abundance of an SLR can fluctuate greatly as a function of Galactic location and time \citep{hotokezaka15, Fujimoto:2018, cote19b, Wehmeyer:2023}. The SLR abundance probability distribution in a parcel of interstellar gas can be evaluated based on the ratio of its mean-life ($\tau=T_{1/2}/\ln{2}$), to the average recurrence time between the formation of its enrichment progenitor ($\gamma$). Larger values of $\tau/\gamma$ equate to a smaller spread in the probability distribution and a greater likelihood that the SLR was present in the ISM material that condensed to form the protosolar molecular cloud. \citet{cote19b} analyses this problem using a statistical framework and concluded that if $\tau/\gamma\gtrsim2$ the relative abundance uncertainty can be quantified and we can be confident that the ESS abundances of these SLRs is due to the buildup of radioactive matter in the ISM. For $\tau/\gamma\lesssim0.3$, instead, it is more likely that the SLR comes from only one event, in which case it is only possible to calculate the time from the last nucleosynthesis event that added radiogenic material to the ESS matter \citep{cote2021}. A further complication is the fact that $\tau/\gamma$ can only be estimated for a given SLR, since the value of $\gamma$ for different stellar enrichment events is poorly constrained. Therefore, a self-consistent origin scenario for all the SLRs in the ESS is complex and remains an ongoing challenge.

The aim of this paper is to analyse the galactic background contribution of $^{53}$Mn and $^{60}$Fe in the ESS. To this end, we use the same GCE framework as in \cite{cote19} and \cite{Trueman:2022} to consider differences in the predicted radioactive-to-stable abundance ratios due to the choice of stellar nucleosynthesis yields, galactic gas inflow histories, and supernova parameterization. For each GCE setup, $t_{\text{iso}}$ is calculated with uncertainties that take into account heterogeneities in the SLR abundances due to stochastic enrichment \citep{cote19b} and the associated error bar on the measured ESS ratios. The derived timescales are compared to those for the neutron-capture SLRs analysed in \cite{Trueman:2022}, to see whether a self-consistent origin scenario can be found in the context of GCE.     

\section{Method}

We use the \texttt{OMEGA+} (One zone Model for the Evolution of GAlaxies) GCE code to predict the radioactive-to-stable abundance ratios of $^{53}$Mn/$^{55}$Mn and $^{60}$Fe/$^{56}$Fe in the Milky Way disk at the time of the formation of the Sun, $t_{\odot}=8.4$ Gyr. Since the identity of SN Ia progenitor remains an open question \citep[see][for a recent review]{Liu:2023}, we consider potential changes in the predicted SLR ratios for different choices of (i) the fraction of SN Ia in the Galaxy that have a sub-Chandrasekhar mass progenitor; (ii) the sub- and near-$M_{\rm{Ch}}$ SN Ia nucleosynthesis yields; and (iii) the distribution of explosion timescales (i.e., delay time distribution) for the population of sub-$M_{\rm{Ch}}$ progenitor. Furthermore, we investigate nucleosynthetic uncertainties in the production of these SLRs by calibrating our GCE model for three different CCSNe yield sets. Finally, we consider also changes in the GCE due to different gas infall timescales for the formation of the Milky Way's thick and thin disk populations. In this Section we give a brief description of the GCE code and its parametrization and our choices of mass- and metallicity-dependent stellar nucleosynthesis yield sets.

\subsection{The \texttt{OMEGA+} GCE framework}

The \texttt{OMEGA+} code is publicly available and is part of the NuGrid Python Chemical Evolution Environment (NuPyCEE) that facilitates the ability to simulate the chemical enrichment and stellar feedback of stellar populations \citep{cote18}\footnote{\url{https://github.com/NuGrid/NUPYCEE}}. The GCE model consists of a central star forming ``galaxy'' region surrounded by a hot gas reservoir that acts as the circumgalactic medium (CGM). Star formation occurs only within the galaxy, which can exchange material with the circumgalactic medium via gas inflows and outflows. The general equation solved by the code at each timestep is 

\begin{equation}
    \delta M_{\text{gas}} = \delta M_{\text{g, in}} + \delta M_{\text{ej}} - \delta M_{\star} - \delta M_{\text{g, out}},
\end{equation}
where $\delta M_{\text{gas}}$ is the change in mass of gas in the central galaxy region due to gas accretion from the CGM ($\delta M_{\text{g, in}}$), gas being returned back to the ISM from dying stars ($\delta M_{\text{ej}})$, gas consumed during the formation of new stars ($\delta M_{\star}$), and feedback driven outflows of gas into the CGM $(\delta M_{\text{g, out}})$.   

In this work, we assume that the Milky Way's thick and thin disks were formed by two distinct gas accretion episodes, such that the rate of inflow at time \textit{t} is described by 

\begin{equation}
    \dot{M}_{\text {g, in }}(t)=A_{1} \exp \left(\frac{-t}{\tau_{1}}\right)+A_{2} \exp \left(\frac{t_{\max }-t}{\tau_{2}}\right),
\end{equation}
where the magnitude of the first and second infall are given by $A_1$ and $A_2$, and the gas accretion timescales are $\tau_1$ and $\tau_2$, respectively. The parameter $t_{\max}$ is the time for maximum accretion of material during the second infall, which is analogous here to the delay time between the maxima of the two infall episodes since we set $A_2=0$ for $t<t_{\max}$. The canonical value for $t_{\max}$ is $\sim1$ Gyr \citep{chiappini:1997, Boissier:1999, Romano:2010}, which is also adopted in this work. However, it has recently been shown that the high- and low-$\alpha$ sequences observed in the observational data of \cite{Silva:2018} can best be reproduced using a ``Revised" two-infall model, which has a significantly longer delay time between first and second infalls \citep{Spitoni:2019}. Furthermore, it was found in \cite{Trueman:2022} that the GCE predicted $^{107}$Pd/$^{182}$Hf ratio at $t_{\rm{iso}}$ was consistent with the isolation times derived individually for $^{107}$Pd ($T_{1/2}=6.5$ Myr) and $^{182}$Hf ($T_{1/2}=8.90$ Myr) only if the former had an additional $9-73\%$ contribution. An enhanced enrichment of $^{107}$Pd relative to $^{182}$Hf may be achieved if high-metallicity low- and intermediate-mass stars contributed to the ESS material; these stars could have formed prior to the birth of the Sun from the gas following a period of prolonged star formation from gas the first gas infall. For these reasons, we also calculate predicted radioactive-to-stable abundance ratios for $^{53}$Mn and $^{60}$Fe with $t_{\max}=4.3$ Gyr as in \cite{Spitoni:2019}. All inflows in this work are assumed to be of primordial metallicity. As in \citet{Spitoni:2019}, galactic outflows are omitted from the Revised infall model. However, in the canonical two infall models where $t_{\max}=1$ Gyr, galactic winds are assumed to be driven by stellar feedback and are proportional to the star formation rate (SFR). The constant of proportionality is the mass-loading factor, $\eta$, which is calibrated so as to remove enough gas from the Galaxy that the observed present-day gas mass of the disk is reached within the uncertainty limits.        

The star formation rate at $t$ is directly proportional to the mass of gas in the central galaxy region, such that  

\begin{equation}
    \dot{M}_{\star}(t) = \frac{\epsilon_{\star}}{\tau_{\star}}M_{\text{gas}}=f_{\star}M_{\text{gas}}(t),
\end{equation}
where $\epsilon_{\star}$ and $\tau_{\star}$ are the star formation efficiency and star formation timescale in the Kennicutt-Schmidt law \citep{Kennicutt:1998}, respectively. For the initial mass function (IMF) we use that of \cite{Kroupa:2013}. At each timestep, the code creates a simple stellar population (SSP) with total mass proportional to the SFR and a distribution of stellar masses in the population according to the IMF. All stars in a given SSP are coeval and have the same initial metallicity, but eject over different timescales according to the lifetimes of the specific stellar nucleosynthesis models used for the yields. The \texttt{SYGMA} \citep[Stellar Yields for Galactic Modelling Applications;][]{ritter:2018} code allows for the contribution toward each species ejected from different stellar sources to be tracked separately in the ISM.

\subsection{Type Ia supernovae}

The galactic SN Ia rate is normalised such that $\sim 1.5\times10^{-3}$ progenitor are born per stellar mass formed in a SSP. It is assumed that both sub- and near-$M_{\rm{Ch}}$ white dwarf (WD) progenitors contribute to this overall rate with a relative frequency of $f_{\rm{sub}}$ and $1-f_{\rm{sub}}$, respectively. Due to the uncertain nature of SN Ia progenitor, we assume $f_{\rm{sub}}$ to be a free parameter from 0 to 1. Predicted radioactive-to-stable abundance ratios are calculated for different combinations of sub- and near-$M_{\rm{Ch}}$ SN Ia yield sets in the literature.

The near-$M_{\rm{Ch}}$ yields are based on a deflagration-to-detonation transition explosion, where the flame front initially propagates through the material at subsonic velocities (deflagration), before transitioning to a full detonation when the burning occurs at supersonic velocities. During the period of deflagration, the fuel is burnt at low enough densities to synthesise the required amounts of intermediate mass elements (e.g., Ne, Mg, Si, S, and Ca) observed in SN Ia spectra. The transition of the flame front to a detonation occurs once the density drops below $\sim10^7\text{g cm}^{-3}$ and it is energetic enough to completely unbind the WD. The yields of \cite{Seitenzahl:2013} are based on the post-processing of the explosion of a $1.40M_{\odot}$ model simulated using a three-dimensional hydrodynamic code. The strength of the deflagration phase is determined by the number of ignition kernels within a fixed central radius, where the model with 1 ignition kernel experiences the weakest degree of pre-expansion of the fuel and the model with 1600 the highest. In this work we use the yields from the model with 100 ignition kernels (the N100 model) and a central density at explosion of $\rho_c=2.9\times10^9\text{g cm}^{-3}$, which is post-processed at metallicites of $Z=0.02, 0.01, 0.002, 0.0002$. 

Another set of near-$M_{\rm{Ch}}$ yields used in this work are from the double detonation model of \cite{keegans:22}, which are calculated by post-processing the $1.40M_{\odot}$ model of \cite{townsley:16} at 13 initial metallicites in the range $Z=0-0.1$. The model is detonated by inserting a hot spot into the WD, which has a central density at ignition of $\rho_c=2\times10\text{g cm}^{-3}$. We also adopt the yields of the benchmark models of \cite{leung2018} based on explosions of white dwarfs with central burning configurations and a C/O ratio of unity. These benchmark models are calculated for a low ($\times10^9 \text{g cm}^{-3}$), medium ($3\times10^9 \text{g cm}^{-3}$), and high ($5\times10^9 \text{g cm}^{-3}$) central density, corresponding to WD masses of $1.33, 1.38, 1.39M_{\odot}$, respectively. We note that the yields from a pure deflagration near-$M_{\rm{Ch}}$ model are not included in this work, since these are the most likely candidates for the sub-luminous class of SN Iax which are not thought to contribute significantly to the GCE of the solar neighbourhood at the time of the birth of the Sun \citep{Kobayashi:2020a}.   

\begin{figure}
    % \centering
    \includegraphics[width=1.0\linewidth]{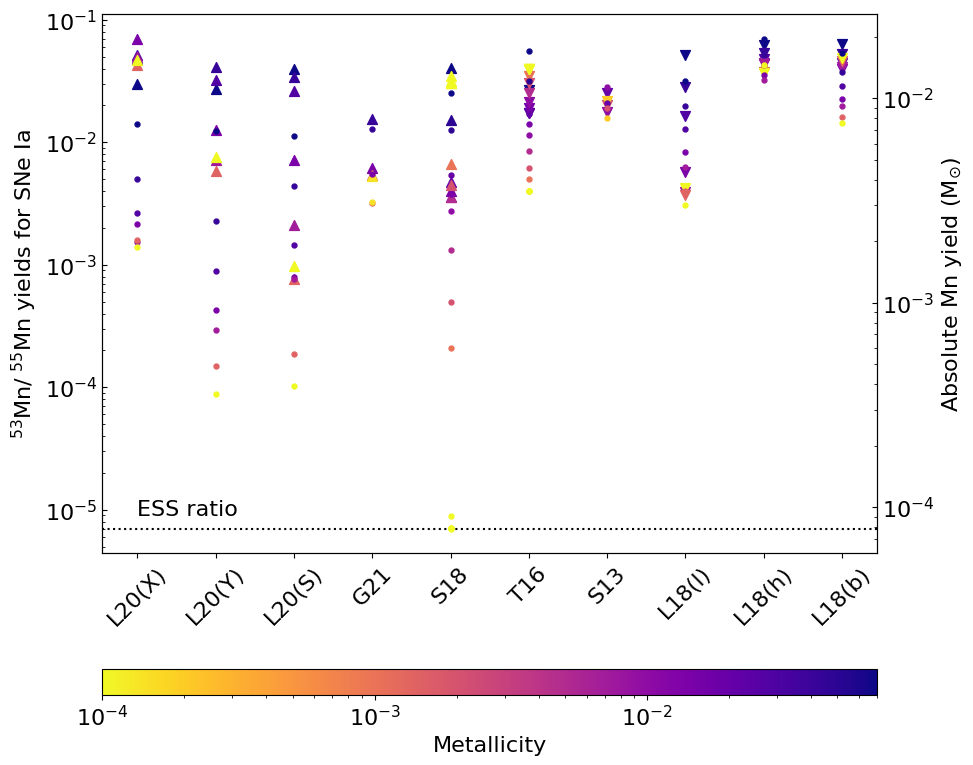} 
     \includegraphics[width=1.0\linewidth]{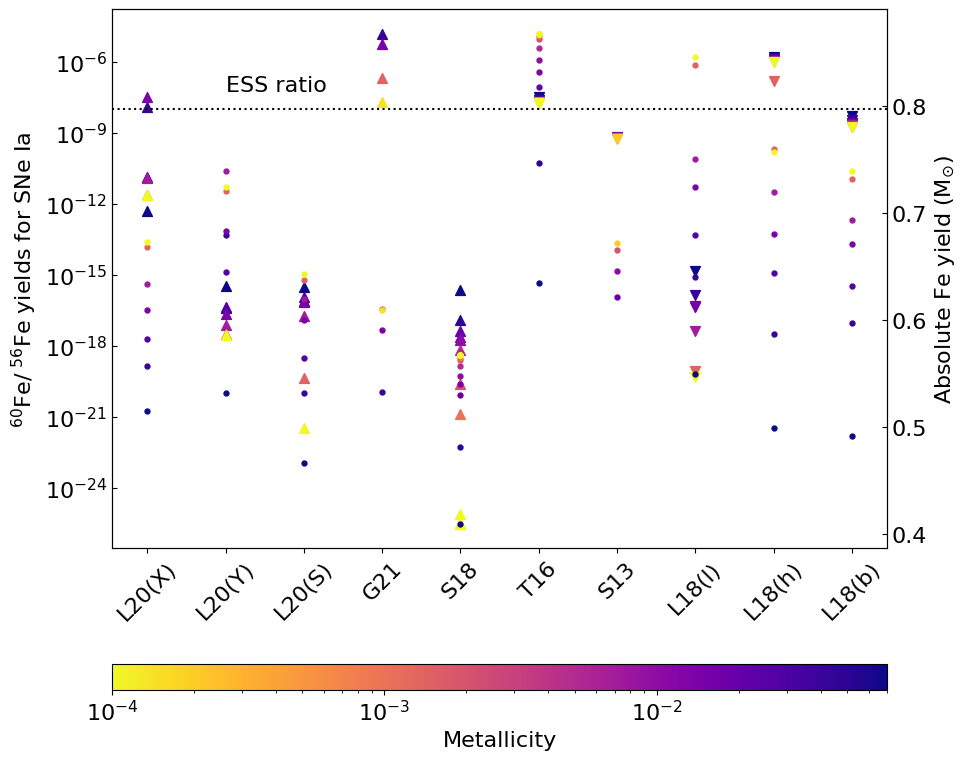}
    \caption{Radioactive-to-stable yields for $^{53}$Mn (top) and $^{60}$Fe (bottom) ejected by the SN Ia models considered in our GCE analysis, and separated by metallicity (different colour). The respective ESS ratios are shown by the dotted lines. Yields for sub- and near-$M_{\text{Ch}}$ models are shown by triangles and upside-down triangles, respectively. The small circles show instead the absolute yields of Mn and Fe ejected by each SN Ia event, as indicated by the secondary y-axes.}
    \label{fig:sn1a_yields}
\end{figure}

For sub-$M_{\rm{Ch}}$ SN Ia, the chosen yields sets are calculated assuming a double-detonation (DD), where the accretion of He-rich material via Roche-lobe overflow onto the WD surface causes a preliminary He-shell detonation that disrupts the WD triggering the secondary central detonation that unbinds the star. In \cite{gronow:21b, gronow:21gce} an hydrodynamic code is used to simulate double detonations resulting from WD progenitor with core masses from $0.8$ to $1.1 M_{\odot}$ and He-shell masses between $0.02$ and $0.1M_{\odot}$. The yields from the model of a $1 M_{\odot}$ WD with a $0.03 M_{\odot}$ He-shell used in this work are calculated by post-processing the explosion at $Z/Z_{\odot}=0.01, 0.1, 1, 3$. The explosion of the $1.0 M_{\odot}$ C/O WD in \cite{shen:2018} was post-processed by \cite{keegans:22} for the same metallicities as the \citet{townsley:16} model described above. Since the explosion is triggered with no He-shell, the model represents a dynamically driven double detonation, where the He detonation is assumed to trigger the C detonation but does not contribute to the overall yields. In \cite{leung:20}, the same code as in \cite{leung2018} is used to model double detonations in sub-$M_{\rm{Ch}}$ WD progenitors taking into account the geometry of the surface He detonation. Therein, yields for the three benchmark models are used, which are based on explosions of WD with masses $1.0, 1.10, 1.10 M_\odot$ and spherical (S), ring (Y), and bubble (X) He detonation geometries, respectively. These models are post-processed at the following metallicities: $Z=0.07, 0.042, 0.028, 0.007, 0.0014, 0$. Prior to the detonation, the S and Y models accrete $0.05M_{\odot}$ of He, whereas the X model accretes $0.1 M_{\odot}$ of material.

Figure \ref{fig:sn1a_yields} compares the radioactive-to-stable yields for $^{53}$Mn (top panel) and $^{60}$Fe (bottom panel) from the SN Ia models considered in our GCE analysis. The absolute yields of Mn and Fe are also plotted for each model, so that it is easier to interpret the overall significance of each enrichment event towards the GCE. For $^{53}$Mn/$^{55}$Mn, all of the SN Ia models, regardless of metallicity, eject a higher radioactive-to-stable ratio than the ESS value. Models with higher initial metallicity tend to produce more $^{53}$Mn relative to $^{55}$Mn than explosions with the same progenitor but at lower metallicity. The near-$M_{\text{Ch}}$ SN Ia produce more total Mn than the sub-$M_{\text{Ch}}$ SN Ia, especially at subsolar metallicities; in the context of the GCE predictions this means that the amount of stable Mn in the ISM at $t_{\odot}$ scales inversely with $f_{\text{sub}}$. For $^{60}$Fe/$^{56}$Fe, there is a much larger range in the yields from SN Ia than for $^{53}$Mn/$^{55}$Mn, although the value of the ratio does not appear to depend on whether the progenitor is sub- or near-$M_{\text{Ch}}$. As for $^{53}$Mn/$^{55}$Mn, relatively more of the SLR is produced compared to the stable reference isotope during burning at higher initial metallicities. Although \cite{gronow:21b}, \cite{townsley:16}, and the high density model of \cite{leung2018} produce $^{60}$Fe/$^{56}$Fe ratios higher than ESS, SN Ia are $\sim10$ times less frequent than CCSNe at solar metallicity, and since the total mass of Fe ejected per event is comparable between these two types of enrichment events, the amount of $^{60}$Fe in the ISM at any given time originates mainly from CCSNe.

The delay time distribution (DTD) for the near-$M_{\rm{Ch}}$ SN Ia population, is assumed to always follow a power law with exponent $-1.07$, which is based on the empirically derived rate of \cite{maoz:12}. The minimum delay time is set at $\sim600$ Myr. A power law DTD is expected for SN Ia originating in a double degenerate system, where the timescale of explosion is dominated by the merger timescale \citep{ruiter:09, toonen:2014}. There is also evidence from empirical studies that supports the existence of a SN Ia population with short delay times of the order of $\sim100-500$ Myr  \citep{Sullivan:2006, Li:2011, Smith:2012}. This so-called ``prompt" delay time can also be retrieved from Binary Population Synthesis (BPS) models, specifically for the case of a binary system in which a sub-$M_{\rm{Ch}}$ mass WD accretes He-rich material from a H-stripped stellar companion \citep{Ruiter:2011}. Since there is no consensus regarding the ratio of the single-degenerate to double-degenerate SN Ia channels, or indeed the value of $f_{\rm{sub}}$, we calculate two suites of GCE models that assume different values of the sub-$M_{\rm{Ch}}$ SN Ia DTD: (1) a tardy population that explodes over the same timescales as the near-$M_{\rm{Ch}}$ population; and (2) a prompt population that follows the DTD derived by \citet{ruiter:14} for a sub-$M_{\rm{Ch}}$ WD accreting He-rich material from a non-degenerate companion, which has a sharp peak in the distribution at $200-300$ Myr.

\subsection{Massive stars and core-collapse supernovae}

We calibrate our GCE framework for three different sets of mass and metallicity dependent massive star yields. Two of the yield sets are from \cite{limongi:18} (hereafter LC18), which include models calculated for masses in the range $10-120 M_{\odot}$ at initial metallicities [Fe/H]$\;=-3,-2,-1,0$. Both sets account for the effects of rotationally induced mixing, averaged as in \citet{prantzos:2018}. Models with initial masses $>25M_{\odot}$ collapse to a black hole and thus the only contribution from these star to the GCE process is via their stellar winds. For the LC18 Set R, the position of the inner border of the mixing and fallback mechanism is chosen such that [Ni/Fe]$\;=0.2$ in the CCSNe ejecta, and the location of the mass cut is such that the ejecta contains $0.07M_{\odot}$ of $^{56}$Ni. For the LC18 Set I, the only constraint is that the supernovae eject $0.07M_{\odot}$ of $^{56}$Ni. The third massive star yield set used in this work is from \cite{nomoto13}, hereafter N13. We do not include hypernovae in this analysis. All GCE models use the low- and intermediate-mass stellar yields from FRUITY \cite{cris08, cristallo09, cris11, cristallo:2015}.

\begin{figure*}
\centering
    \includegraphics[width=.9 \linewidth]{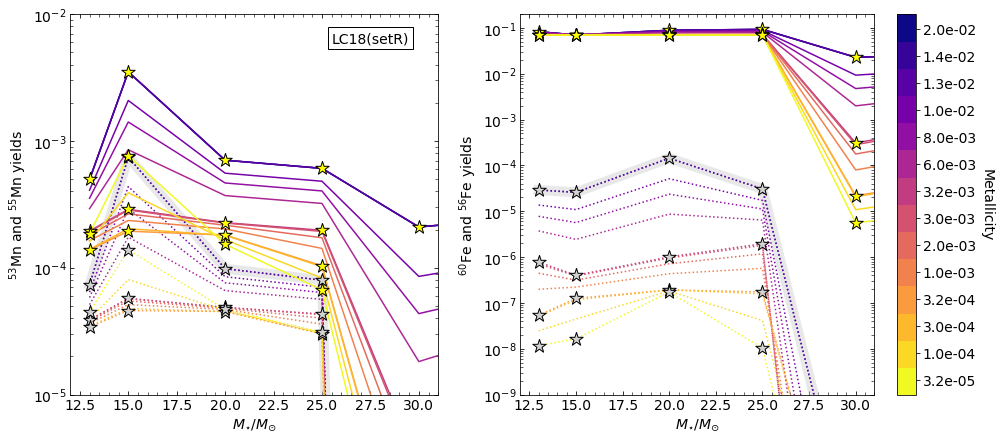}
    
    \includegraphics[width=.9 \linewidth]{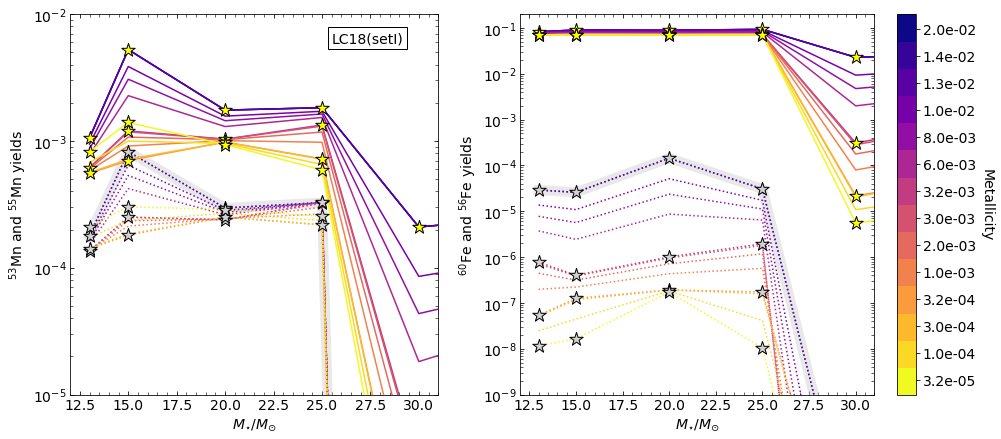}

     \includegraphics[width=.9 \linewidth]{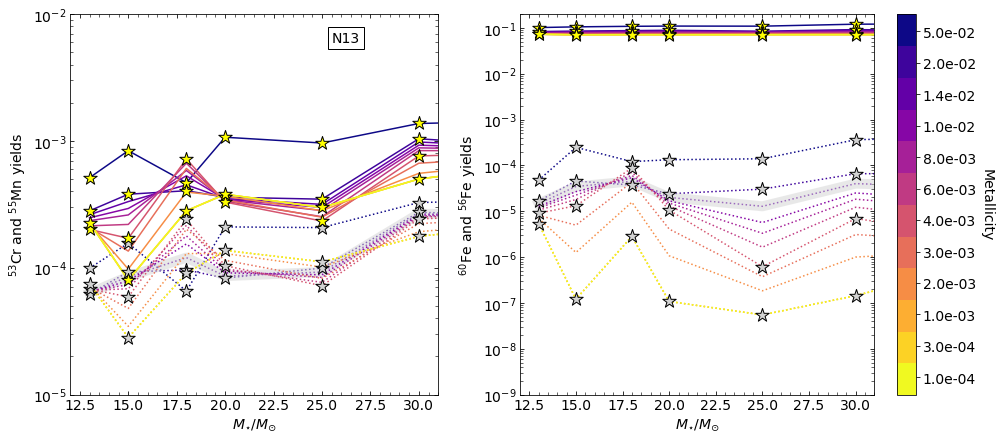}
    
\caption{Total ejected yields in solar masses of $^{53}$Mn and $^{60}$Fe (dotted lines) and their respective stable reference isotopes (solid lines) for the massive star (wind + CCSN) nucleosynthesis yield sets used in this work. The top and middle panels are the Set R and Set I yields from \cite{limongi:18} respectively, and the bottom row are the yields from \cite{nomoto13}. The latter set does not include $^{53}$Mn, so we use the abundance of its stable daughter isotope, $^{53}$Cr, as its proxy. Stars indicate models for which yields are calculated, and the lines are obtained by their interpolation. Thick grey lines indicate SLR yields for massive stars with $Z=Z\odot$. For the \cite{limongi:18} sets, only the rotationally averaged yields for a given mass and metallicity are used, as calculated in \cite{prantzos:2018}.}
\label{fig:production_ratios}
\end{figure*}

Figure \ref{fig:production_ratios} shows the CCSN yields for the SLRs $^{53}$Mn and $^{60}$Fe and their respective stable reference isotopes, $^{55}$Mn and $^{60}$Fe, for each of the three yield sets used in our analysis. The yields show the summation of both the pre-explosive and explosive nucleosynthetic components (i.e., winds + CCSNe), however, the last will dominate since the Fe-peak elements are mostly synthesised at high temperatures and densities in the mist internal parts of the ejecta \citep[e.g.][]{Woosley:1995, Thielemann:1996, woosley:2002}. Whilst the SLR contribution to the solar material from sub-solar metallicity stars is negligible at the time of the formation of the Sun, they contribute meaningfully toward the accumulated buildup of the stable reference isotopes. Stars with solar-like metallicities are most relevant for the SLR abundances in the ESS, so a thick grey line is used to highlight the SLR yields for $Z=Z_\odot$ in Figure \ref{fig:production_ratios}. Since the $M_{\star}/M_{\odot}\geq25$ LC18 models are assumed to collapse into black holes and thus do not eject CCSN material, the Set R and Set I models eject substantially less of the SLRs (and their reference isotopes) of interest compared to their N13 counterparts in this mass range. In general, the $^{53}$Mn/$^{55}$Mn ratio decreases at higher metallicities because more $^{55}$Mn is present in the initial material of the progenitor star, and the production of freshly synthesised Mn during incomplete Si-burning decreases for larger $Y_{e}$ \citep{nomoto13}. The production of $^{60}$Fe depends on the $^{59}$Fe branching ratio -- that is, the ratio between the $^{59}$Fe(n,$\gamma$)$^{60}$Fe reaction and the $^{59}$Fe $\beta$-decay rates \citep{limongi06, Jones:2019:60Fe, Yan:2021, Spyrou:2024} -- and therefore is favoured at higher neutron densities and at temperatures $<2$ GK. There is a noticeable shift in the yield trends of $^{53}$Mn and $^{55}$Mn for the LC18 yields at $15M_{\odot}$ due to a merger between the silicon-burning shell and oxygen-rich layers in some of the models with rotational velocities of $150$ and $300$ Km s$^{-1}$; this merger disrupts the relative Fe-peak abundance ratios in the CCSN ejecta, most notably for the case where the Ni/Fe ratio is fixed in Set R.

\subsection{GCE model calibration}

\begin{figure*}
    \centering
    \includegraphics[width=.4 \linewidth]{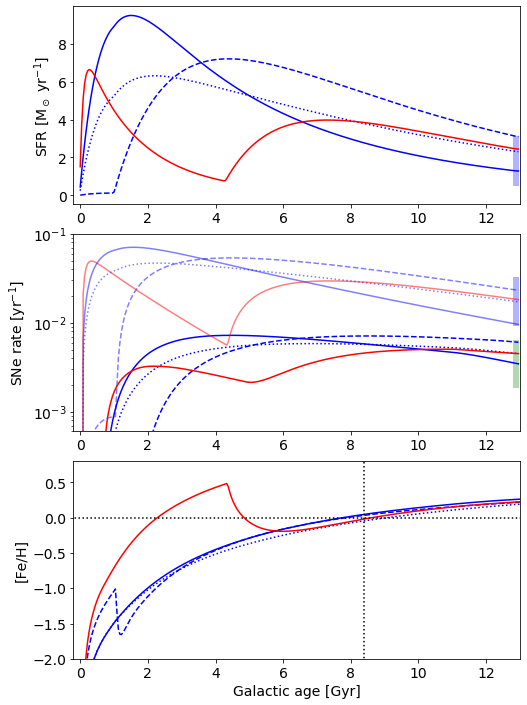} 
    \includegraphics[width=.394\linewidth]{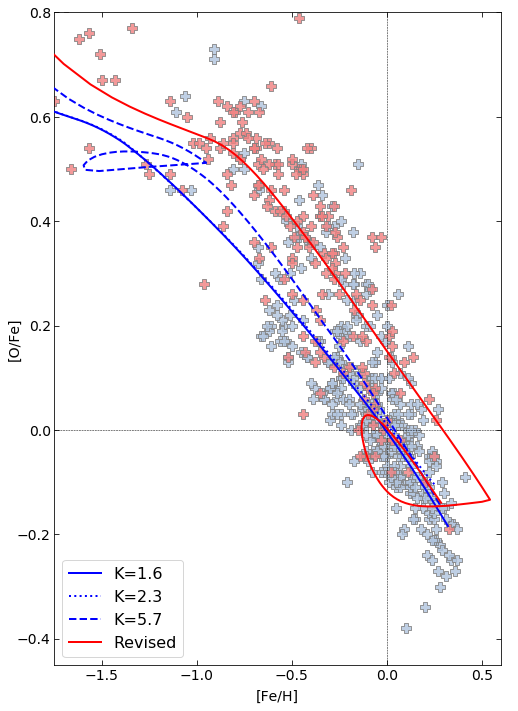} 
    
    \caption{Star formation rate (top left panel), supernova rates (middle left panel), global [Fe/H] evolution (bottom left panel), and [O/Fe] versus [Fe/H] (right panel) for the four GCE setups. The GCE models are calculated using a $50\%$ contribution from sub- and near-$M_{\rm{Ch}}$ progenitor. Blue lines are the GCE setups with canonical $t_{\max}=1.0$ Gyr, whereas the red line has $t_{\max}=4.3$ Gyr. The vertical bands at Galactic time of $13$ Gyr indicate the present day observed rates from \cite{kubryk:2015}, where the blue and green bands in the middle panel are for CCSN and SN Ia, respectively. The horizontal and vertical dotted lines in the bottom panel indicate solar [Fe/H] and $t_{\odot}$, respectively. Observational data in the right panel are thick (red) and thin (grey) disk data from \cite{bensby:2014}.}
    \label{fig:fig_GCE_calib}
\end{figure*}

The \texttt{OMEGA+} framework is used to calculate four different GCE setups. Three of the setups are based on the canonical two-infall model of \cite{chiappini:2001} with $t_{\text{max}}=1$ Gyr, where each corresponds to a different value of the free parameter $K$ in the analytical solution for the galactic inflow rate defined in \cite{clayton84, clayton88} as  

\begin{equation}
    \dot{M}_{\text{inflow}}(t)=\frac{K}{t+\delta}M_{\text{gas}}(t).
\end{equation}
These GCE setups are calibrated against several observational constraints to deliver an upper, a lower, and a best estimate for the radioactive-to-stable abundance ratios in the ISM at $t_{\odot}$. The observations used are present-day estimates for the SFR, supernova rates, inflow rates, and total mass of gas in the Milky Way's disk. Three of them are equivalent to the `low', `best' and `high' setups in \cite{cote19} and \cite{Trueman:2022}, described by $K=1.6$, 2.3, and 5.7, respectively. The fourth model is the Revised infall model of \cite{Spitoni:2019}. The left panels of Figure \ref{fig:fig_GCE_calib} show the evolution of the SFR, supernova rates, and [Fe/H] for the four different GCE setups. The GCE models in the figure are all run with $f_{\rm{sub}}=0.5$, a tardy SN Ia DTD, and using the \cite{leung2018} and \cite{leung:20} yields for the near- and sub-$M_{\rm{Ch}}$ SN Ia, respectively. It is not necessary to calibrate each GCE setup individually for all yield combinations and values of $f_{\rm{sub}}$, since the SN Ia yields all eject very similar amounts of Fe. In the top panel, there is a clear decrease in the SFR during the period between the first and second infall in the Revised model; this is to be expected since in \texttt{OMEGA+} the SFR is proportional to the mass of gas in the galaxy region and, since no new gas is being added, it is quickly consumed by star formation. In the middle panel, the CCSN rates of the models (faded lines) very closely follow the same profile as the star formation rate, since the delay time between progenitor formation and eventual explosion is several orders of magnitude shorter than the galactic timescale. However, this is not the case for the SN Ia rates which causes a noticeable delay between the peaks in the SN Ia rate and the SFR. Finally, in the bottom panel the global [Fe/H] of the ISM is compared to the solar value for the four models. The [Fe/H] rises quickly for the Revised model since there is no outflows to remove metal enhanced gas. At the time of the start of the second infall ($t_{\max}=4.3$ Gyr) in the Revised model the metallicity is supersolar ([Fe/H]$\;\sim0.5$), but sharply decreases once primordial gas is reintroduced into the system. Since there is no appreciable gap in the star formation for the non-Revised model, the [Fe/H] evolves smoothly with Galactic age. All models are also calibrated to reach the solar [Fe/H] at $t_{\odot}=8.6$ Gyr.   

Figure \ref{fig:fig_GCE_calib} (right panel) shows the [O/Fe] evolution as a function of metallicity for each of the GCE setups compared to thick and thin disk data from \cite{bensby:2014}. Both the high and Revised tracks have a characteristic loop feature, which is a consequence of the gas metallicity decreasing  following the accretion of fresh primordial material during the second gas inflow episode \citep[see also, e.g.,][]{Spitoni:2019, palla:2021}. This feature is not seen in the low and best models because the magnitude of the second inflow parameter is small enough that the accretion of metal-poor gas never dominates over the net increase in metallicity of the gas due to stellar enrichment. In general, all models reproduce well the observational data. In particular, the Revised model can better reproduce the majority of the high-$\alpha$ data at low metallicities in the thick disk, but also predicts that some amount of low-[O/Fe] stars with [Fe/H]$\;\approx0.5$ should be observed, for which only very few can be accounted for in the sample.

\section{Results}

In this section we analyse the radioactive-to-stable abundance ratios recovered from our GCE parameter sweep for the SLRs $^{53}$Mn and $^{60}$Fe, and we calculate isolation times for these ratios taking into account uncertainties due to ISM heterogeneities of the SLR abundance and the error associated with the ESS ratios derived from meteorites. We then compare these results to the isolation times for the $s$-process SLRs $^{107}$Pd and $^{182}$Hf calculated in previous work and discuss the \iso{60}Fe/\iso{53}Mn ratio. 

\begin{figure*}
\centering
    \includegraphics[width=.45 \linewidth]{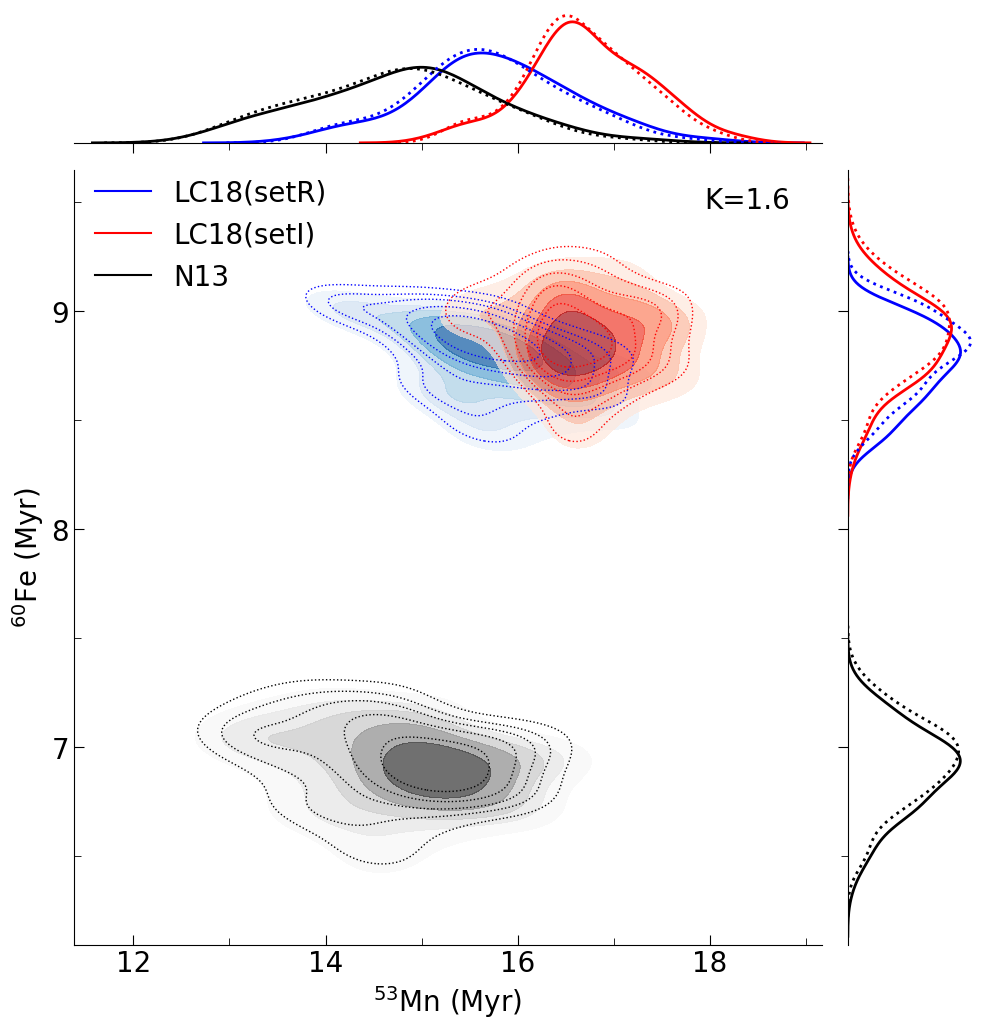}
    \includegraphics[width=.45 \linewidth]{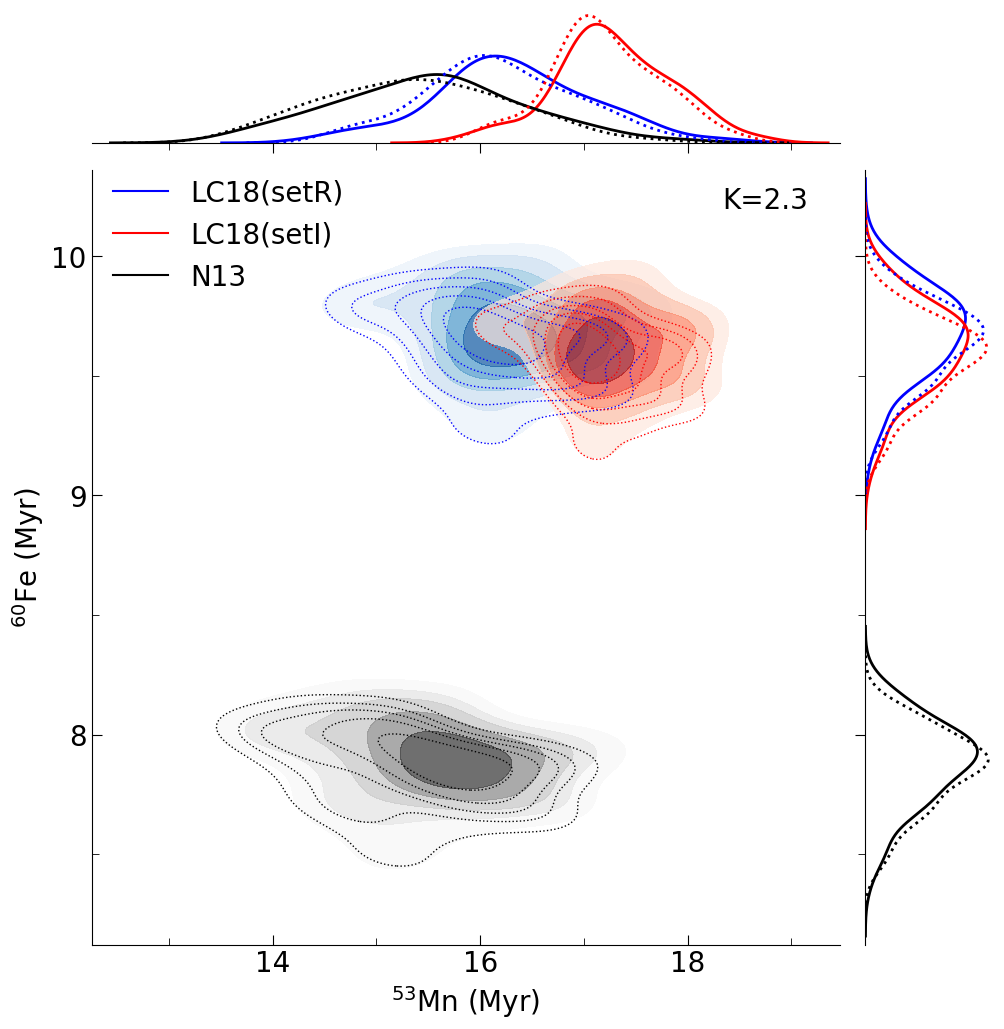}

    \includegraphics[width=.45 \linewidth]{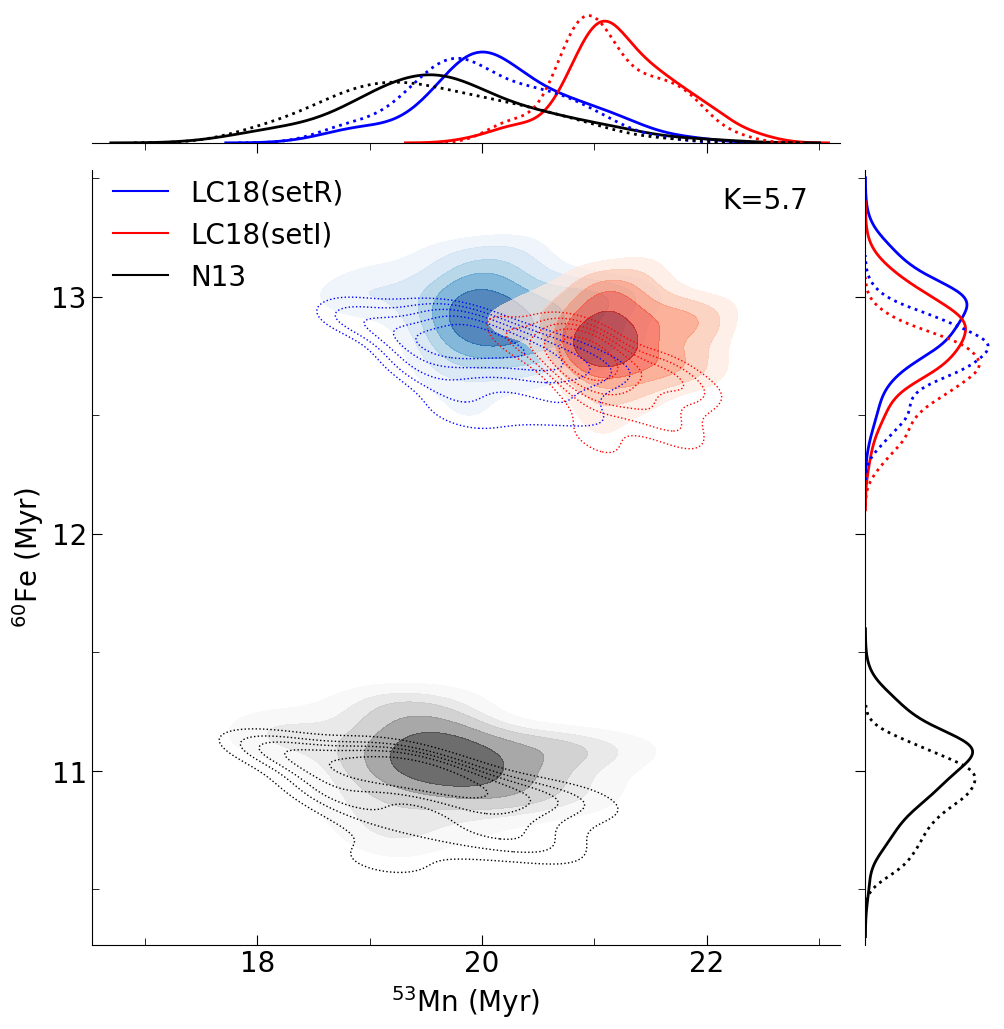}
    \includegraphics[width=.45 \linewidth]{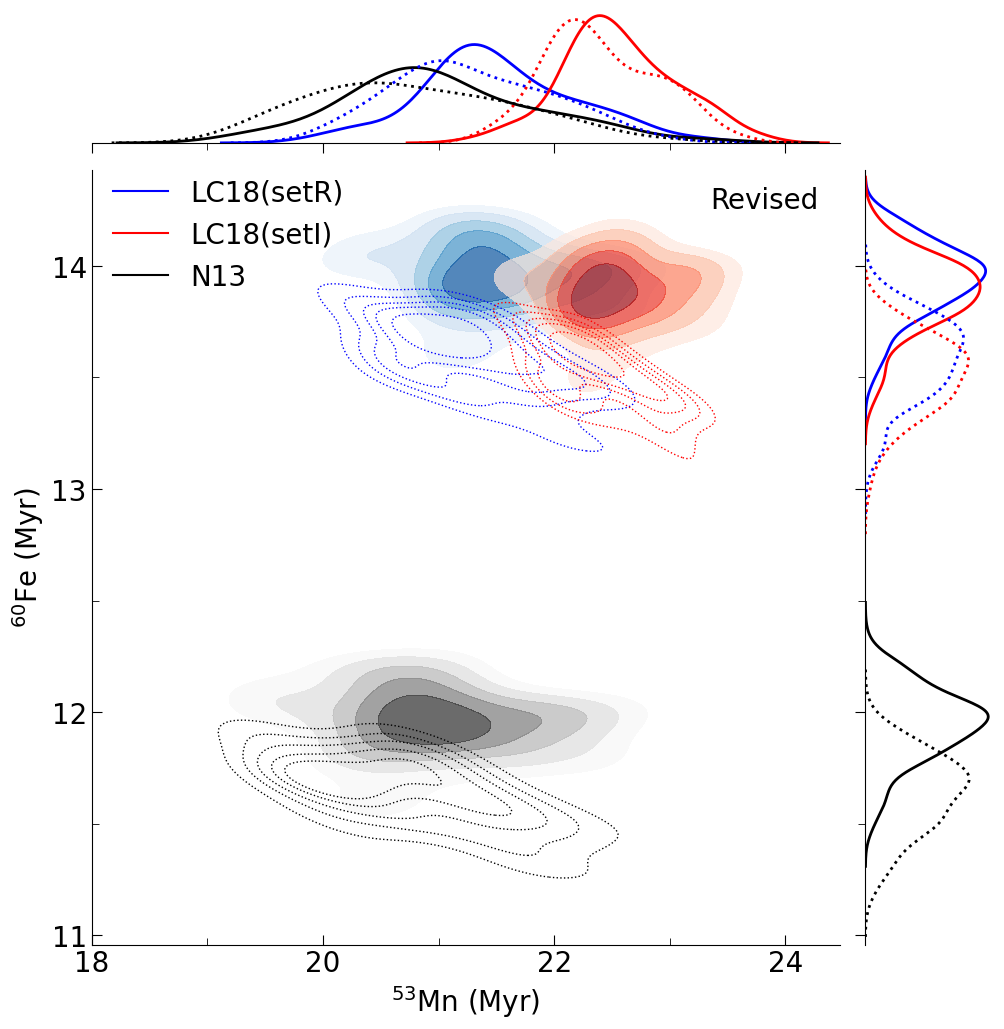}
\caption{Comparison of $^{53}$Mn and $^{60}$Fe isolation times derived using the predicted SLR-to-stable ratios in the ISM at $t_{\odot}=8.4$ Gyr for each of the four different GCE setups. Each panel shows separately the times derived for GCE models that use the massive star yields from  \cite{limongi:18} Set R (blue), Set I (red), and \cite{nomoto13} (black). The kernel density estimations (KDEs) for GCE setups that use the prompt DTD for the sub-$M_{\text{Ch}}$ population have a dotted outline. Each KDE takes into account the spread of times predicted by GCE models with different SN Ia yield combinations and values of $f_{\text{sub}}$. Contours are plotted at $20\%$ intervals.}
\label{fig:fig_kde_plots}
\end{figure*}

\subsection{Isolation times for $^{53}$Mn and $^{60}$Fe}

The isolation time is the time taken for the GCE predicted radioactive-to-stable abundance ratios at $t_{\odot}$ to decay to the ESS ratios inferred from meteoritic analysis. Figure \ref{fig:fig_kde_plots} shows kernel density estimations (KDEs) calculated for the $^{53}$Mn (x-axis) and $^{60}$Fe (y-axis) isolation times derived from the GCE models at $t_{\odot}=8.4$ Gy, plotted as separate panels for each of the four GCE setups of Figure \ref{fig:fig_GCE_calib}. Six separate distributions are plotted in each panel, which correspond to models calculated with one of the three different CCSN yield sets (differentiated by colour) and for each of the tardy and prompt sub-$M_{\text{Ch}}$ SN Ia DTDs (differentiated by a dotted outline for the prompt distributions). Within a given panel, each KDE shows the spread of isolation times that are derived for a given GCE setup -- that is, the same star formation efficiency, gas inflow rate, etc. -- with different combinations of the SN Ia yields and relative sub- and near-$M_{\text{Ch}}$ fractions. Note that the axis range is not consistent between panels, since each GCE setup represents a unique possible Milky Way evolution, therefore, isolation times between the four setups (panels) must not be compared to each other. The isolation times plotted here do not take into account the uncertainty factors on SLR abundances associated to stochastic stellar enrichment.   

For all CCSN yield sets, the $^{60}$Fe isolation times are shorter than for $^{53}$Mn. The N13 yield set leads to $^{60}$Fe isolation times that are $\sim2$ Myr lower than those from either LC18 yield sets. Referring to Figure \ref{fig:production_ratios}, this discrepancy can be explained by considering that for the CCSN yields at and around solar metallicity \citep[i.e, $Z_{\odot}=0.014$ for][]{asplund:2009} the N13 models eject less $^{60}$Fe than their LC18 counterparts; this results in lower $^{60}$Fe/$^{56}$Fe ratios at $t_{\odot}$ and hence less decay time is required to reach the ESS value. The $^{53}$Mn isolation time distribution for Set I is shifted to $\sim1$ Myr longer times than for Set R, however, there is no discernible difference for $^{60}$Fe which is to be expected since the Fe yields for these CCSN models appear analogous in Figure \ref{fig:production_ratios}. For $K=1.6$ and $K=2.3$, there are negligible differences in delay times for the GCE models that adopt either the prompt or tardy DTD.

As the value of $K$ increases, so to do the $^{53}$Mn and $^{60}$Fe isolation times. We remind that larger values of $K$ correspond to GCE configurations with higher radioactive-to-stable abundance ratios in the ISM at $t_{\odot}$, so this correlation is to be expected. The $^{60}$Fe isolation times for the tardy and prompt distributions also become more distinguishable as $K$ increases, with a difference of $\sim0.5$ Myr for $K=5.7$. Since SN Ia do not efficiently produce $^{60}$Fe, this offset is due to the difference in time delay with which they begin to significantly contribute Fe in the Galaxy. Higher values of $K$ increase the relative contribution of the second infall episode toward the total mass of gas in the Milky Way disk which, in combination with a population of prompt SN Ia, leads to a faster [Fe/H] evolution of the gas at earlier galactic times and thus a greater relative contribution of Fe from SN Ia at $t_{\odot}$. The results for the Revised infall model in Figure \ref{fig:fig_kde_plots} most closely resembles that of the $K=5.7$ model, but with even longer isolation times and more separation for prompt and tardy distributions.              
  
\subsection{Comparison to $^{107}$Pd and $^{182}$Hf isolation times} \label{sec:comparison}

In \cite{Trueman:2022}, isolation times were derived for $^{107}$Pd, $^{182}$Hf, and $^{135}$Cs produced by asymptotic giant branch (AGB) stars for the $K=1.6$, 2.3, and 5.7 GCE setups \footnote{The SLR $^{135}$Cs will not be considered further here because the GCE models in \cite{Trueman:2022} adopted $\tau=3.3$ Myr for this SLR, instead of the more recent value of 1.9 Myr. Furthermore, the AGB yields used in the GCE models were calculated without the inclusion of the new T-dependence of the half-life of \iso{134}Cs \citep{li21}, the branching point leading to the production of $^{135}$Cs. In addition, only an upper limit is available for the ESS $^{135}$Cs/$^{133}$Cs ratios. For further discussion regarding this SLR, we refer the reader to \citet{Leckenby:2024}.} These authors found that in each GCE setup the isolation times were consistent with each other within $1\sigma$ when accounting for uncertainty factors on the SLR abundance ratios due to stochastic chemical evolution. Since the solar isolation time should be consistent for all SLRs that are assumed to have $\tau/\gamma\geq2$, we can compare the results of this previous work to those of $^{53}$Mn and $^{60}$Fe to test for consistency between SLRs produced by very different types of stellar enrichment events. Therefore, in order to consider heterogeneities in SLR abundances in the ISM due to stochastic stellar enrichment, it is necessary to apply the uncertainty factors in \cite{cote19b} to the isolation time distributions in Figure \ref{fig:fig_kde_plots}. 

We remind that the uncertainty factors for a given SLR abundance in a parcel of gas are uniquely defined for each value of the $\tau/\gamma$ ratio. Only for $\tau/\gamma\geq2$, can we be confident that the SLR was inherited by the early Solar System from the ISM material from which it condensed. However, since the value of $\gamma$ is poorly defined for each type of stellar enrichment event, as in \cite{Trueman:2022}, we consider the largest value of $\gamma$ for each SLR for which $\tau/\gamma\geq2$ is true; this gives us the most conservative errors on the isolation times (these uncertainties were calculated for the simplified case where both the SLR and stable reference isotope come from one type of event). If more than one event is significant, as in the case of \iso{53}Mn and \iso{60}Fe, then the total uncertainty will be greater. In our case, given that the two possible sources (CCSNe and SN Iae) are independent from each other, it is enough to consider the sum of the two random variables. The effect of the summation would be to, almost exclusively, increase the GCE uncertainty for the SLR/stable ratio towards larger values, which would translate into longer $t_{\rm iso}$.

To partly address this problem, we also tested a case where we remove all $^{53}$ Mn from the yields of SN Ia, so that \iso{53} Mn in the ESS would originate from only one source. This approach corresponds to removing the underlying assumption that $\tau/\gamma\geq2$ is applicable for the case of $^{53}$Mn enrichment from SN Ia, which is entirely feasible given that SN Ia events are rare relative to the $\tau$ of $^{53}$Mn; and assuming the last SN Ia event to have contributed to the solar material occurred long enough ago prior to the formation of the Sun that the $^{53}$Mn had time to significantly decay. For example, the $^{53}Mn$ produced by an SN Ia enrichment event that occurred $>$50 Myr before the formation of the Sun would decrease by more than a factor of 10,000 prior to being incorporated into the first solids.

\begin{figure*}[h!]
\centering
    \includegraphics[width=.32 \linewidth]{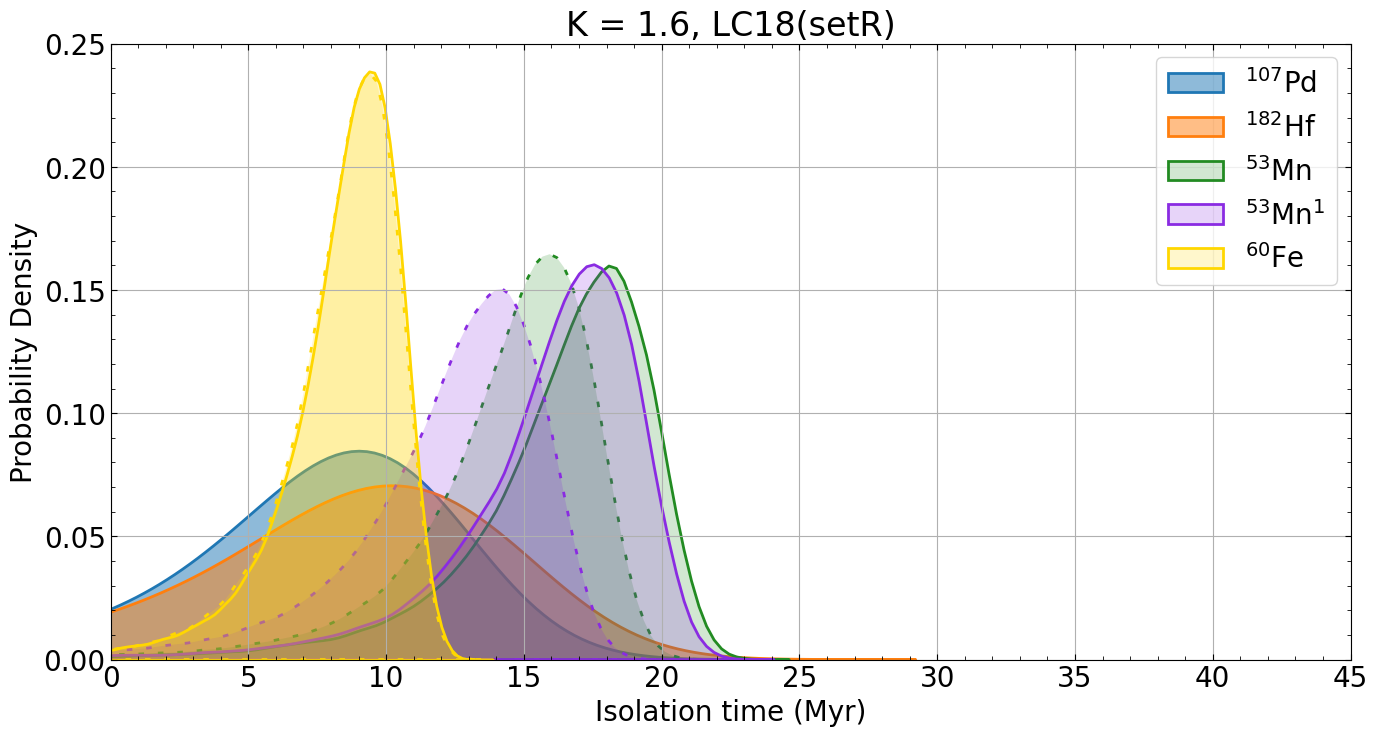}
    \includegraphics[width=.32 \linewidth]{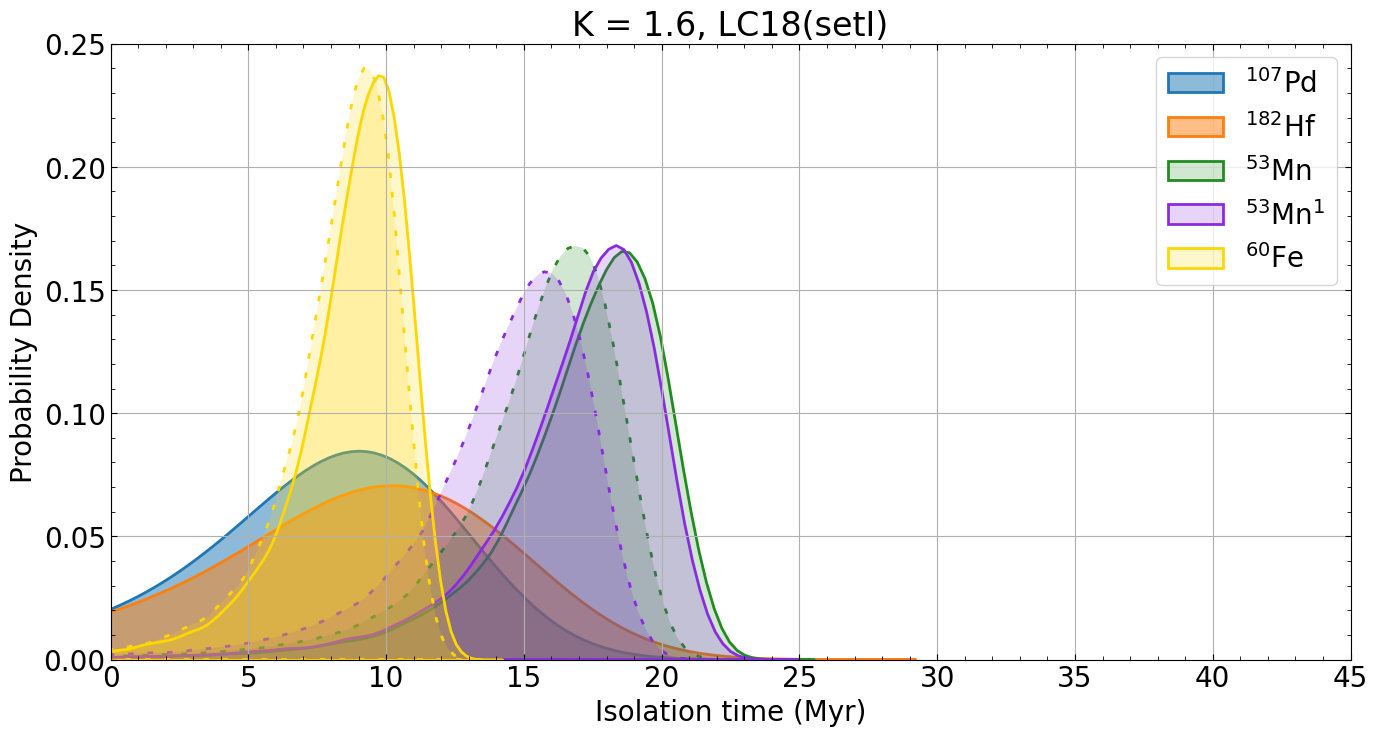}
    \includegraphics[width=.32 \linewidth]{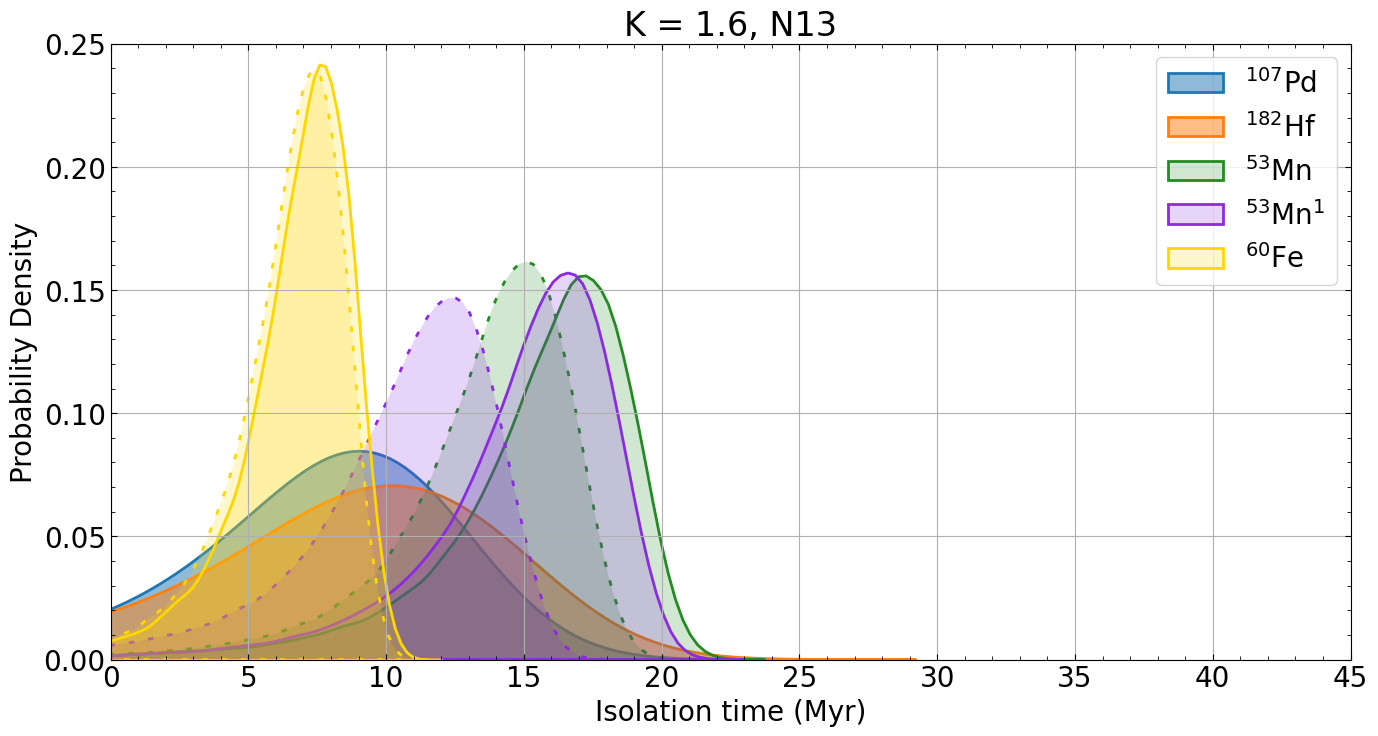}

    \includegraphics[width=.32 \linewidth]{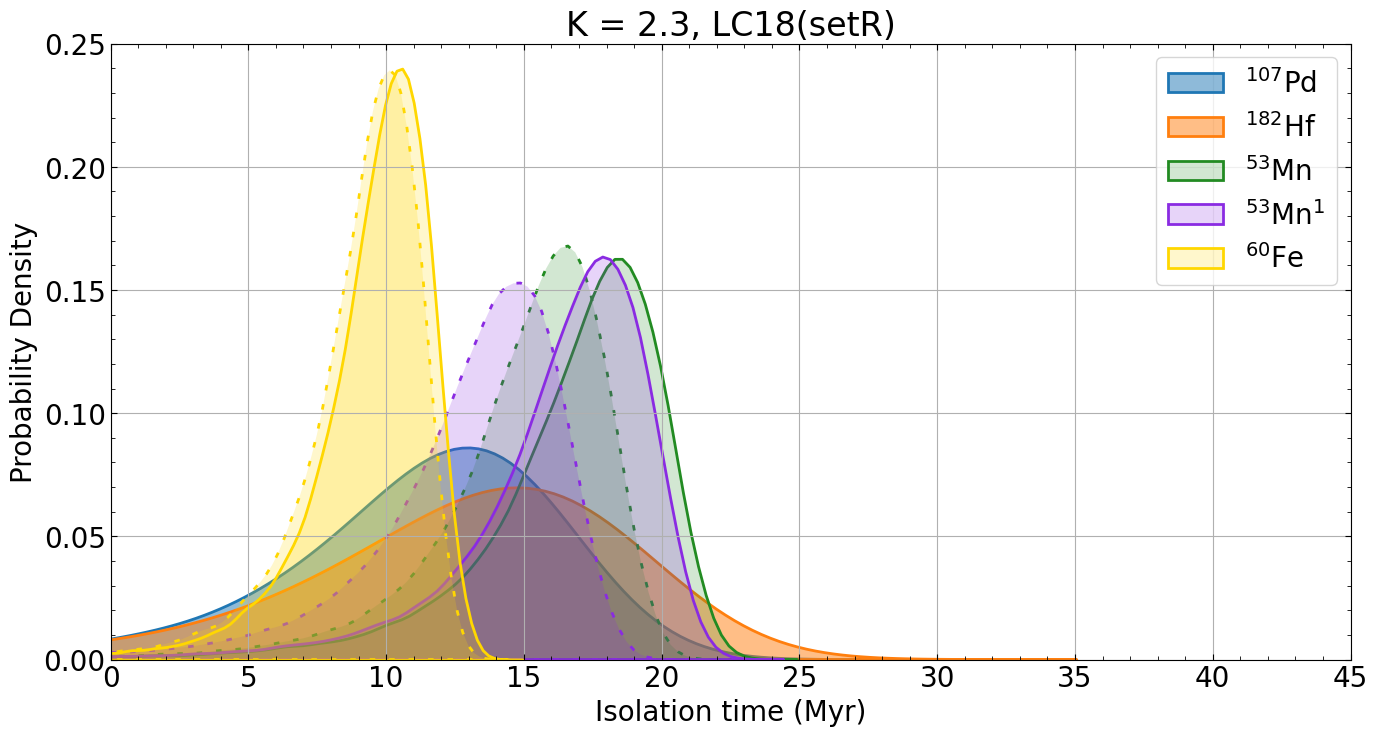}
    \includegraphics[width=.32 \linewidth]{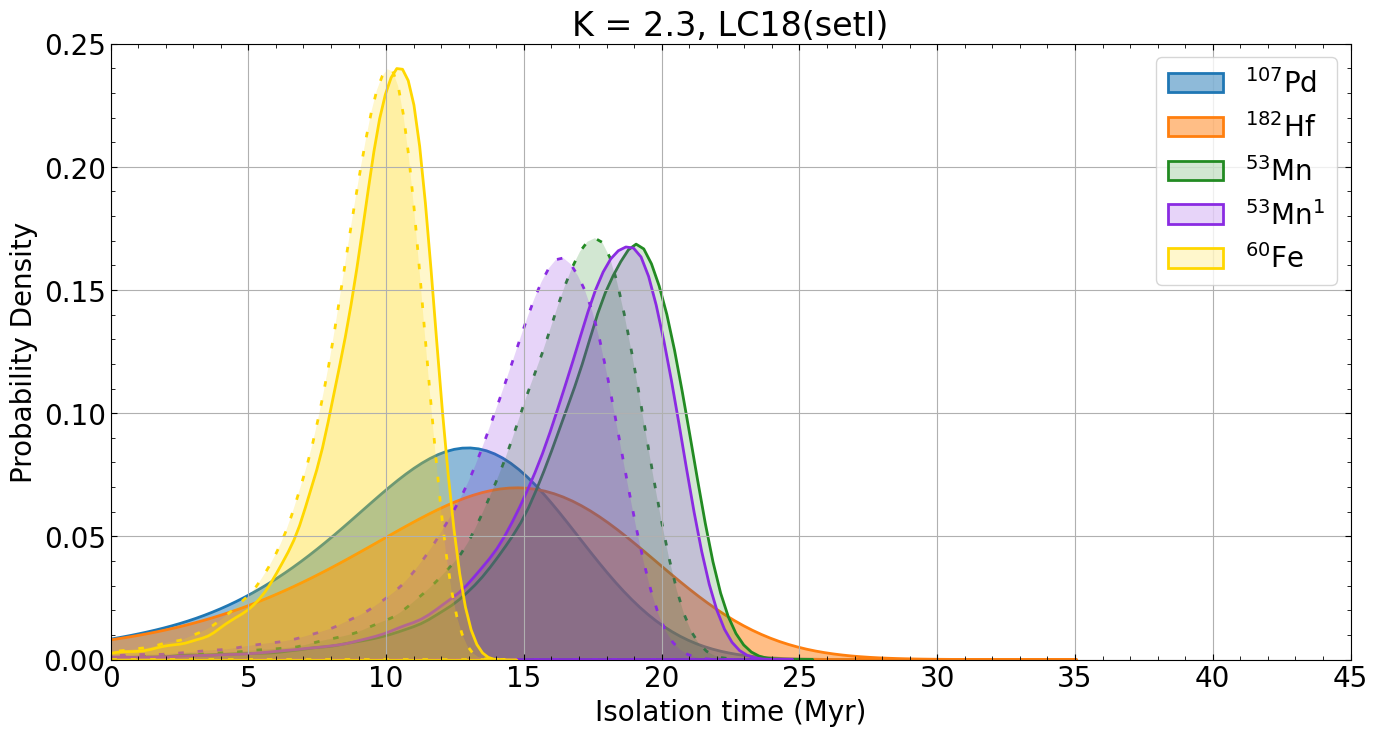}
    \includegraphics[width=.32 \linewidth]{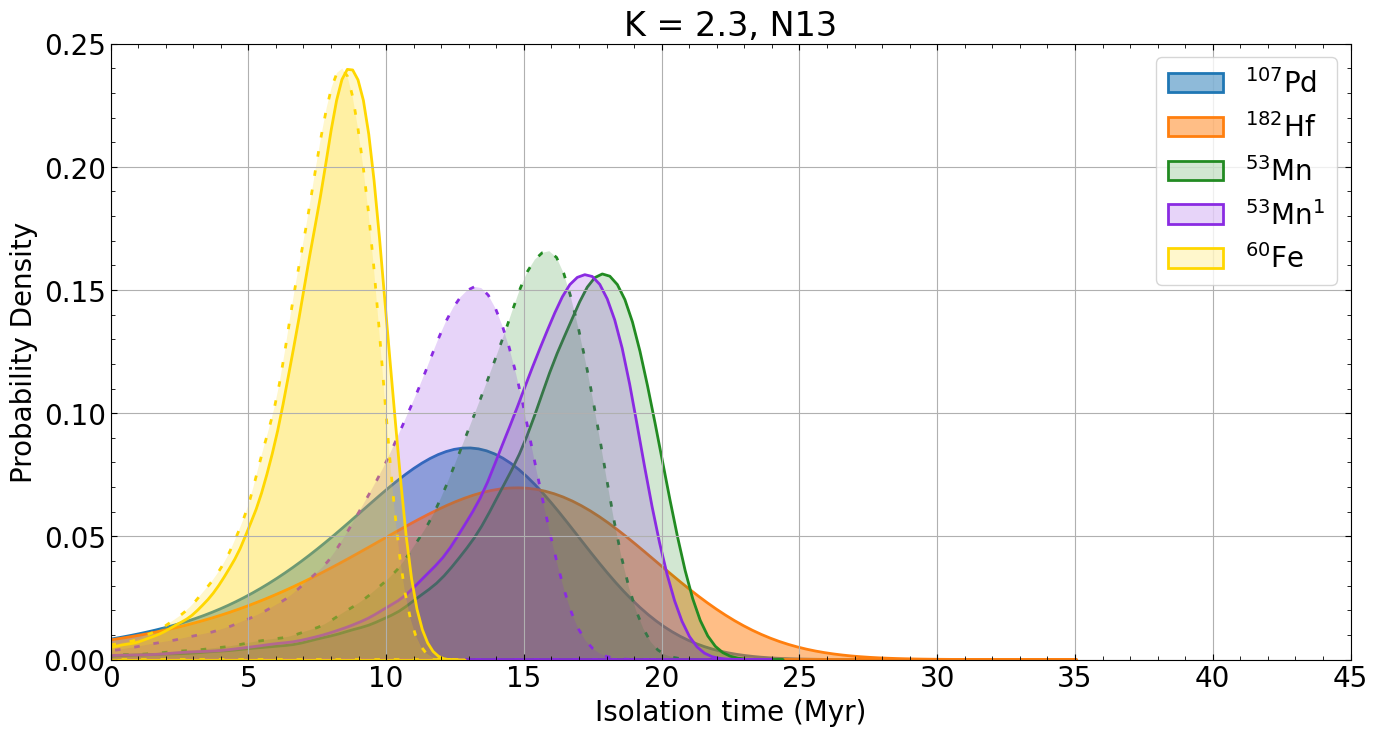}

    \includegraphics[width=.32 \linewidth]{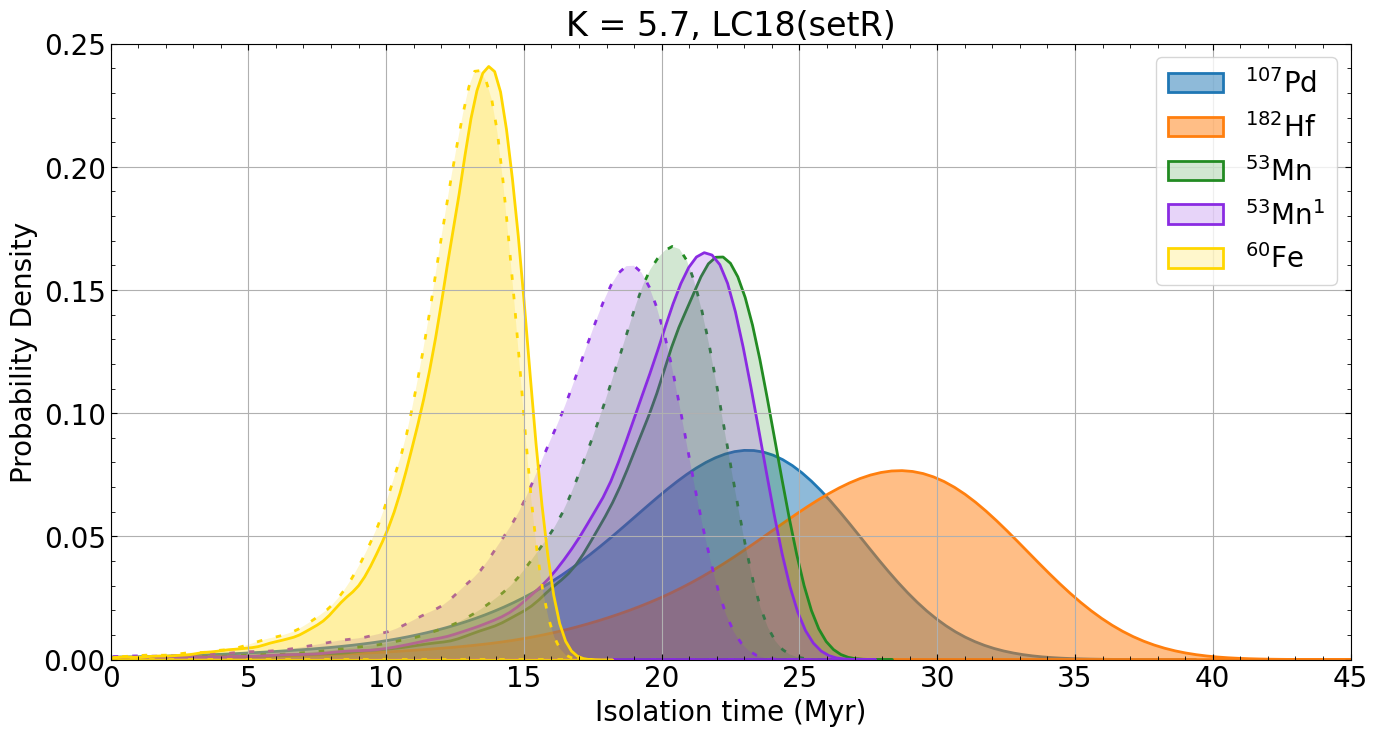}
    \includegraphics[width=.32 \linewidth]{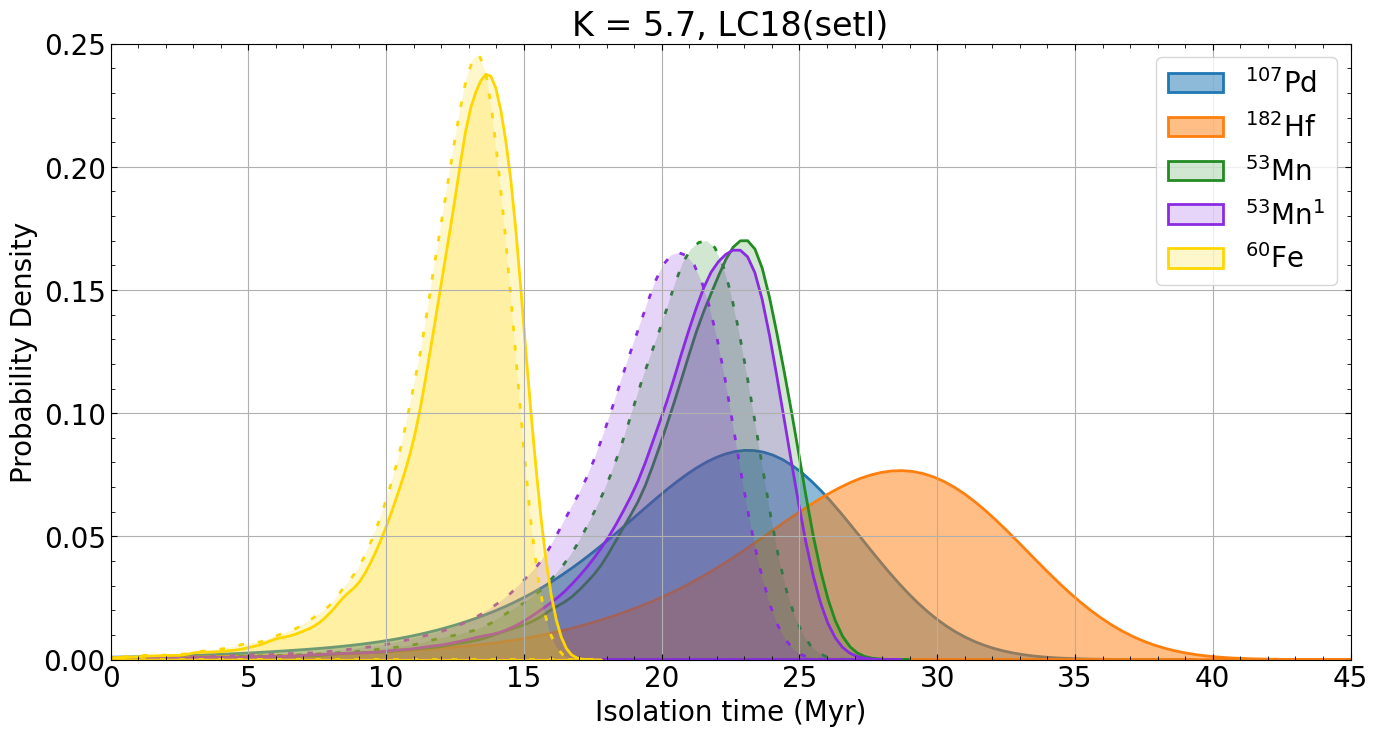}
    \includegraphics[width=.32 \linewidth]
    {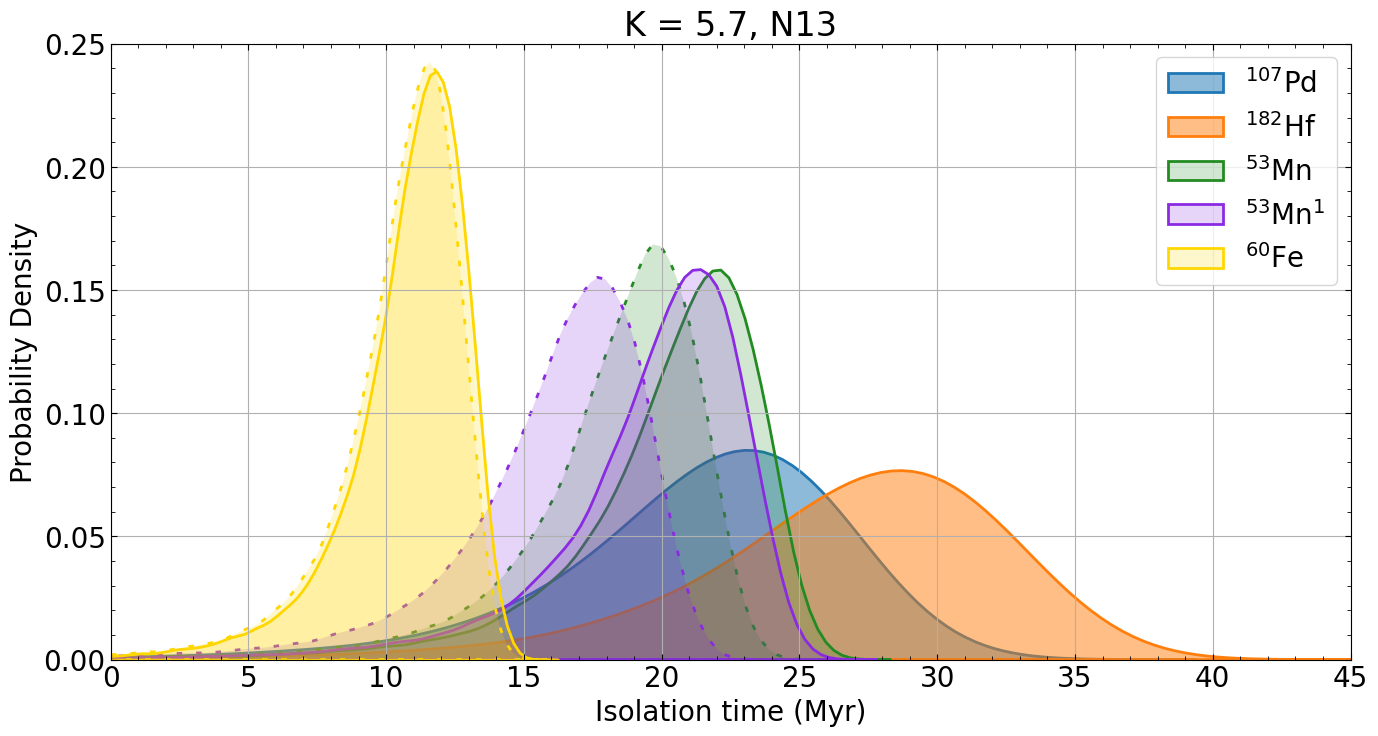}
    
    \includegraphics[width=.32 \linewidth]{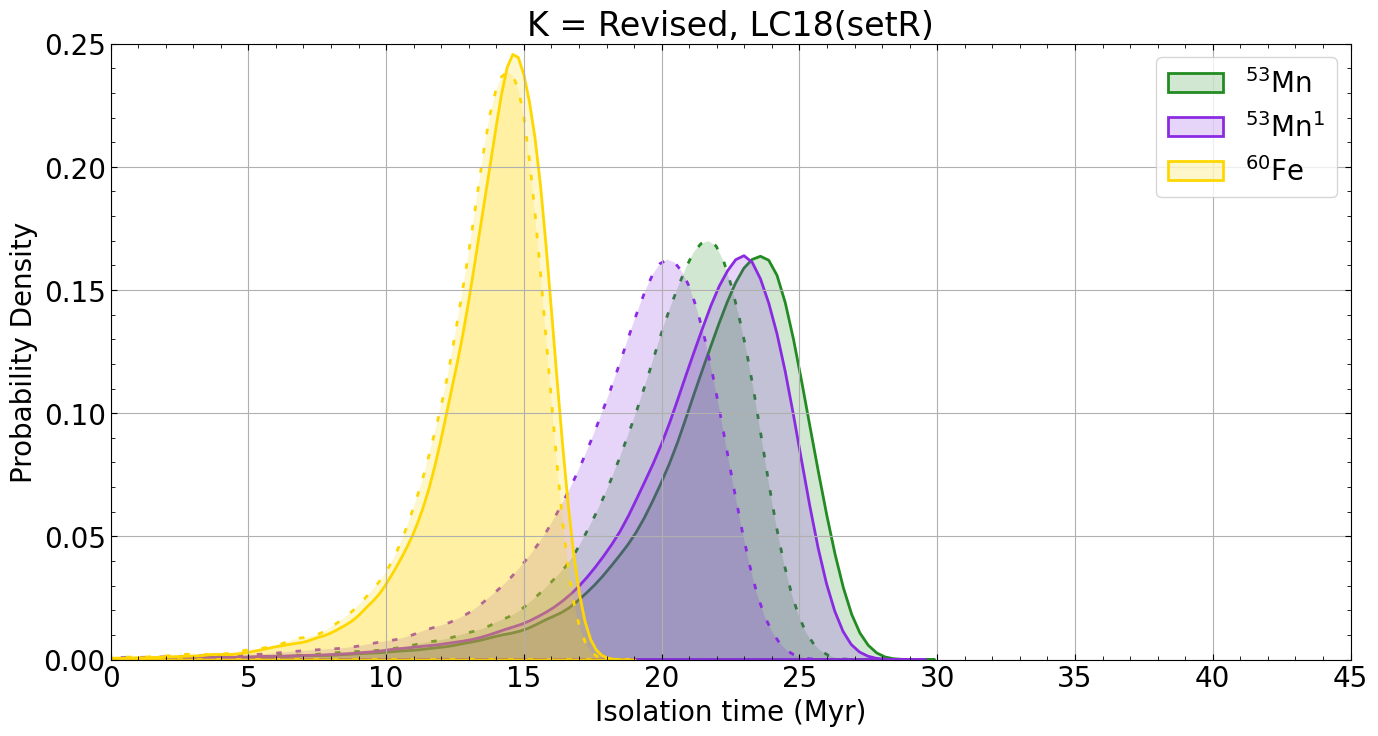}
    \includegraphics[width=.32 \linewidth]{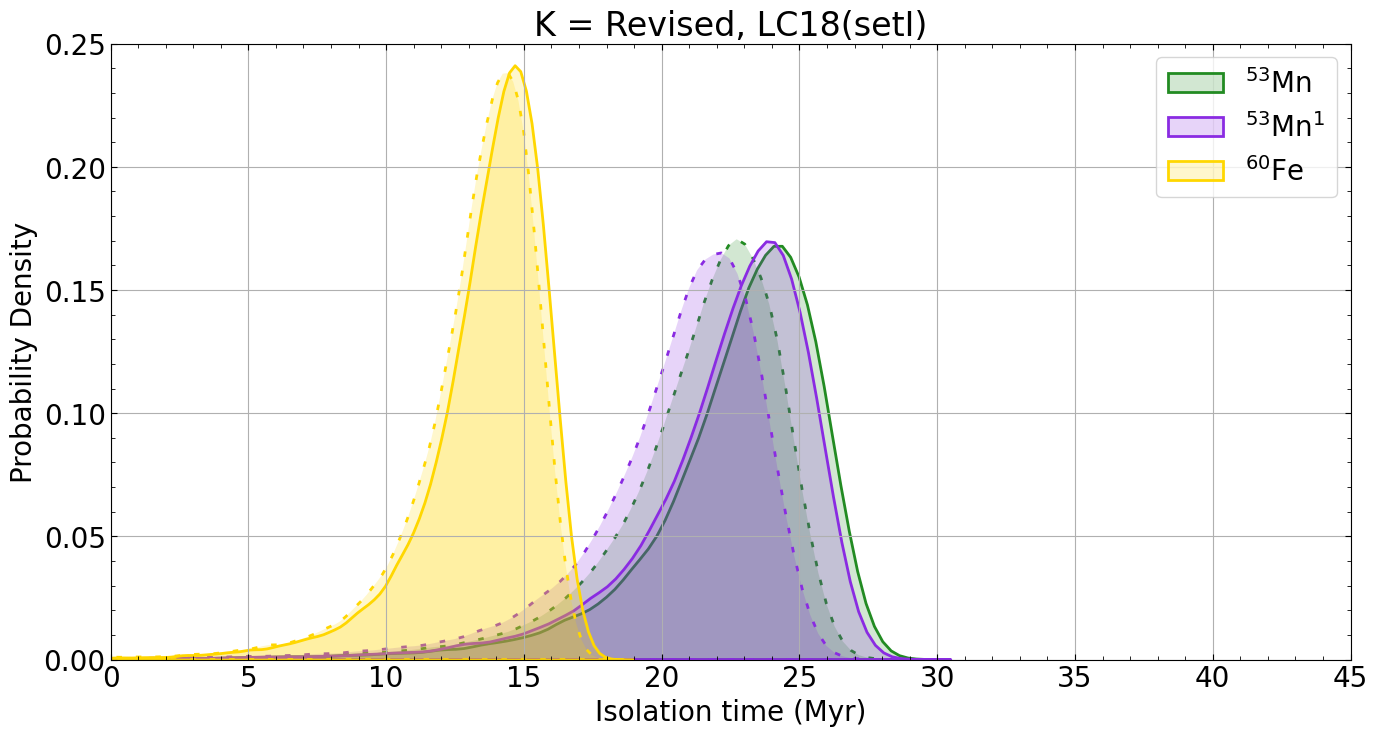}
    \includegraphics[width=.32 \linewidth]
    {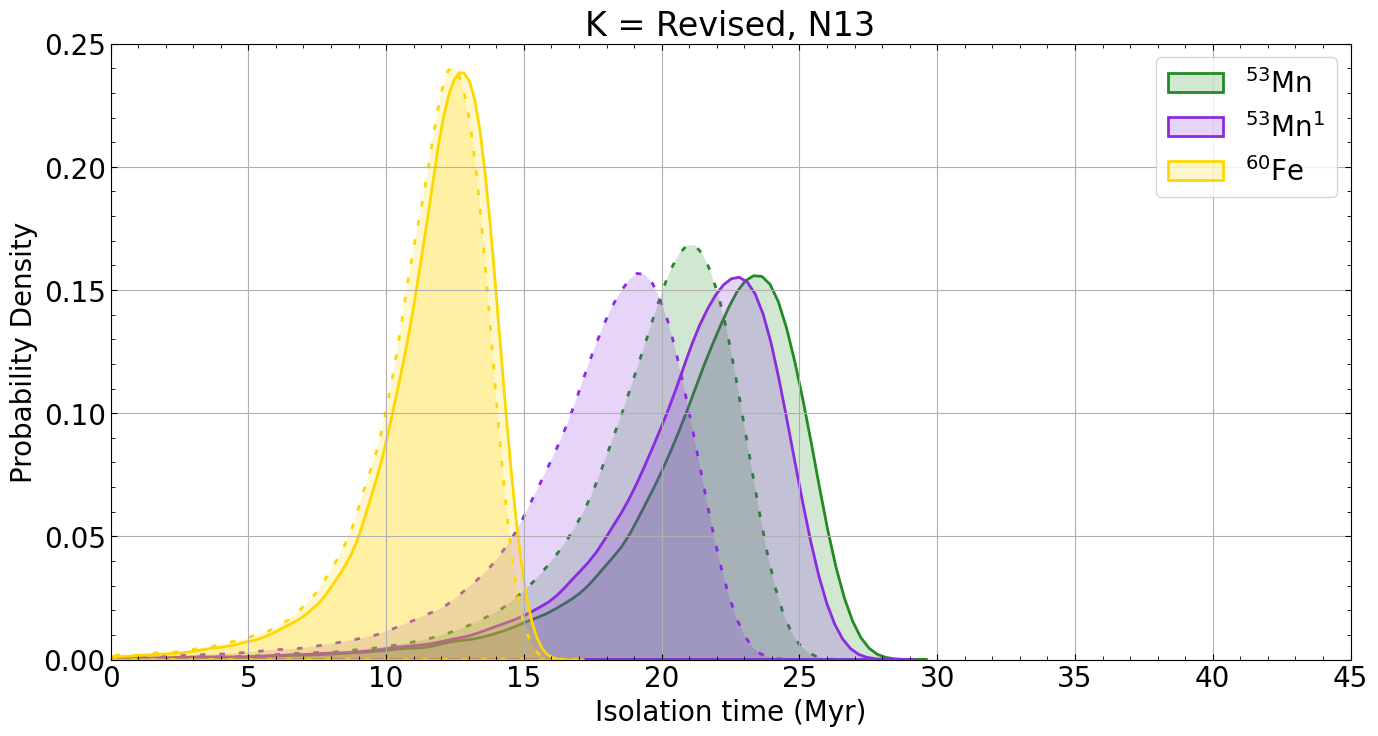}
    
\caption{Comparison of the probability density estimates for the GCE predicted isolation times of $^{107}$Pd and $^{182}$Hf \citep[from][]{Trueman:2022} to those from $^{53}$Mn and $^{60}$Fe (this work) assuming a tardy SN Ia DTD. As indicated at the top of each panel, the four rows correspond to the three values of $K$ and Revised GCE setup, while the three columns correspond to the three different choices of the CCSN yields. For $^{53}$Mn and $^{60}$Fe, results using two different values of $f_{\text{sub}}$ are shown separately, where the solid line represents the maximum value of $f_{\text{sub}}=1.0$ and the loosely dotted line the minimum value of $f_{\text{sub}}=0.0$. The distributions with $0<f_{\text{sub}}<1$ are omitted for clarity, but always lie between the two plotted distributions. The purple distributions labeled as $^{53}$Mn$^1$ are derived for GCE models in which SN Ia are assumed to not contribute any $^{53}$Mn in the ESS. The distributions take into account the uncertainty factors in the SLR abundances due to stochastic chemical enrichment \citep[from][]{cote19b} as well as the uncertainties on the ESS radioactive-to-stable abundance ratios inferred from meteorites.}
\label{fig:fig_norm_dist_tardy}
\end{figure*}

In addition to the uncertainty factors on the GCE predicted radioactive-to-stable abundance ratios, we consider also the uncertainties associated with the ESS ratios inferred from meteorites. The error uncertainty around the mean ESS abundance ratio is assumed to be normally distributed. The uncertainty factors for the GCE predicted ratios can be well modelled assuming a skewed normal distribution centered around the mean, with shape parameter calculated from the $84^{\text{th}}$ percentile spread of the SLR abundance distribution calculated in \cite{cote19b}. By sampling the distributions for the ESS and GCE ratios and applying these to the results in Figure \ref{fig:fig_kde_plots}, a probability density of isolation times can be derived individually for each isotope \citep[see also][]{Leckenby:2024} that takes into account uncertainties due to (i) the precision of the measured value for the ESS ratio, (ii) ISM heterogeneities in SLR abundances due to stochastic chemical enrichment, and (iii) different stellar nucleosynthesis yields. The results of this analysis are shown in Figure \ref{fig:fig_norm_dist_tardy} for the GCE models with the tardy sub-$M_{\text{Ch}}$ SN Ia DTD and in Figure \ref{fig:fig_norm_dist_prompt} for the GCE models with instead the prompt DTD -- the panels are divided by CCSN yield set (separate columns) and by $K$ (separate rows). The blue and red kernel density estimations (KDEs) are for the neutron capture isotopes $^{107}$Pd and $^{182}$Hf as derived in \cite{Trueman:2022} for the Monash AGB yields, whereas the green and yellow distributions are for $^{53}$Mn and $^{60}$Fe (this work). In addition, the GCE setups that include no contribution of $^{53}$Mn from SN Ia are plotted in purple. For \iso{53}Mn and \iso{60}Fe, the KDEs calculated with $f_{\text{sub}}=0$ and 1 are shown separately in each panel to show the most extreme cases, however, the KDEs almost indistinguishable for $^{60}$Fe. In \cite{cote19b} it was found that the standard deviation of the average SLR abundance in the ISM depends on the DTD of the stellar source, with shorter DTDs leading to narrower spreads. It follows that the KDE for $^{60}$Fe is the most narrow since the SLR comes from only one source and the DTD for CCSNe is typically much shorter than for AGB stars and SN Ia, and the uncertainty in the meteoritic ratio is better constrained than for the other SLRs.

For $K=1.6$, there is generally good agreement for the isolation times of $^{107}$Pd, $^{182}$Hf and $^{60}$Fe for all CCSN yield sets. The $^{53}$Mn isolation times are mostly too long compared to those of the other SLRs, especially for $f_{\text{sub}}=1$ (solid outline). Although sub- and near-$M_{\text{Ch}}$ SNe Ia have similar $^{53}$Mn/$^{55}$Mn ratios in their ejecta, the absolute yield of Mn is lower for sub-$M_{\text{Ch}}$ (see Fig. \ref{fig:sn1a_yields}), so the ISM contains less of the stable reference isotope, $^{55}$Mn, at $t_{\text{iso}}$, thus requiring longer isolation times to reach the ESS radioactive-to-stable ratio. The GCE runs with no $^{53}$Mn from SN Ia have shorter isolation times, consequently they have more overlap with the spread of times derived for the other SLRs. The difference between tardy (Figure \ref{fig:fig_norm_dist_tardy}) and prompt (Figure \ref{fig:fig_norm_dist_prompt}) DTDs for the sub-$M_{\text{Ch}}$ SN Ia population is negligible\footnote{In fact, this is true for all GCE setups and yield combinations, so further comparison between the two DTDs is unnecessary.}.

The GCE setups with $K=2.3$ have the highest likelihood for a self-consistent solution since they have the most cumulative overlap between the different distributions. Since the distributions associated with the neutron-capture SLRs are relatively wide, the time intervals are in agreement with both the $^{53}$Mn and $^{60}$Fe distributions; this is in spite of there being only a relatively small probability window where the $^{60}$Fe and $^{53}$Mn distributions overlap each other. In particular, for the CCSN yields of LC18 (setR) and N13 the GCE models with lower $f_{\text{sub}}$ values or with no SN Ia contribution to $^{53}$Mn agree well with the isolation times of $^{182}$Hf. Depending on the CCSN yield set, the peak of the $^{60}$Fe distribution is $\sim3-5$ Myr lower than the most likely isolation times for the neutron-capture SLRs. The largest difference in the maximum likelihoods of the $^{53}$Mn and $^{60}$Fe distributions is for LC18 (setI), however, setR has the most overlap when all distributions are considered.  

For $K=5.7$, only the highest derived isolation times for $^{60}$Fe that have relatively low likelihood are in agreement with the lowest possible isolation times derived for the $s$-process SLRs. For $^{53}$Mn, there is excellent agreement with the $^{107}$Pd distribution, particularly for the GCE models with $f_{\text{sub}} \sim 1.0$. Only isolation times below the median value of the $^{182}$Hf distribution (i.e., $\lesssim29$ Myr) are compatible with those derived for any of the other SLRs. 

In \cite{Trueman:2022}, isolation times for $^{107}$Pd and $^{182}$Hf were not calculated using a Revised GCE setup, therefore KDEs for these SLRs are omitted from the final row of Figs \ref{fig:fig_norm_dist_tardy} \& \ref{fig:fig_norm_dist_prompt}. For $^{53}$Mn and $^{60}$Fe, the KDEs have more narrow distributions than for the other GCE setups, and the Revised setup also has the smallest difference between the distributions with different values of $f_{\text{sub}}$. There is only a small probability window where the $^{60}$Fe and $^{53}$Mn isolation times overlap and, given the shift of the $^{107}$Pd and $^{182}$Hf distributions to longer isolation times for larger values of $K$, we predict that the Revised setup has the lowest likelihood for a self-consistent solution.  

\subsection{Isolation times from $^{60}$Fe$/^{53}$Mn}

In principle, the ratio of the two SLRs, $^{60}$Fe/$^{53}$Mn, can also be used independently to measure the isolation time. This ratio decays with a $\tau$ equivalent of 12.6 Myr, where $\tau_{\rm eq} = \tau_{\rm 53} \cdot \tau_{\rm 60}/(\tau_{\rm 53} - \tau_{\rm 60})$. The ESS ratios of the SLRs is $0.127_{-0.038}^{+0.045}$, calculated using the values of $^{60}$Fe/$^{56}$Fe, $^{53}$Mn/$^{55}$Mn, and the solar abundances of $^{56}$Fe and $^{55}$Mn from \citet{lodders20}. The error bars (given at 2$\sigma$) are found from Monte Carlo (MC) sampling the normal distribution given by the individual error bars associated with the quantities listed above\footnote{In the same way, we calculated the error bar for the $^{107}$Pd/$^{182}$Hf ratio investigated in \citet{Trueman:2022}. The result is an ESS ratio of $4.29_{-0.48}^{+0.54}$ at 2$\sigma$, which does not significantly change the conclusions reported in that paper.}. The GCE MC statistical uncertainty on the ratio of two SLR was analysed and calculated in detail in \citet{yague21}, although the $^{60}$Fe/$^{53}$Mn ratio was not discussed specifically in that work because only isotopes with the same stellar origin were considered. Nevertheless, if we consider the GCE model with the $^{53}$Mn from SNe Ia set to zero, it is possible to derive a stochastic uncertainty with GCE also for this ratio, and use it as a third constraint. However, we did not proceed with this method because 90\% of the GCE models predict ISM $^{60}$Fe/$^{53}$Mn ratios at the time of the formation of the Sun lower than the ESS value (i.e., negative isolation times), while the other 10\% resulted in isolation times lower than 1 Myr, in disagreement with all the values derived so far. 

These results apply also to the GCE setups calculated with no contribution to $^{53}$Mn from SNe Ia, therefore, the predicted $^{60}$Fe/$^{53}$Mn ISM ratios reflect a problem already present in the CCSNe models. Within CCSNe, the $^{60}$Fe/$^{53}$Mn ratio records the relative contribution of different nucleosynthesis processes: $^{53}$Mn is mostly produced by explosive O-burning and partial Si-burning, while $^{60}$Fe is produced in the outer layers of the star via neutron captures in the C- and He-burning regions \citep[e.g.][and references therein]{Lawson:2022}. Therefore, a possibility to increase the $^{60}$Fe/$^{53}$Mn would be provided, at least qualitatively, by decreasing the amount of mass from the inner part of the ejecta that escapes collapsing back onto the compact object. A lower amount of the ejecta from the inner layers would lead to a lower amount of radioactive \iso{56}Ni, corresponding to the so-called weak or faint CCSNe, which are more difficult to observe. The need for a larger contribution of faint CCSNe to the GCE of $^{60}$Fe/$^{53}$Mn as found in this work, would be in agreement with independent models of the evolution of planet-forming elements \citep{pignatari23}, especially their relative abundances, and with the observed scatter of $r$-process elements with respect to Fe from metal-poor stars in the early Galaxy \citep{wehmeyer:2019}. Our work would further support the idea that faint CCSNe play a significant role in GCE.

\section{Conclusion}

We evaluated the potential origin of the SLRs $^{53}$Mn and $^{60}$Fe in the ESS using full GCE models that account for the production of these isotopes and their reference stable isotopes, $^{53}$Mn and $^{56}$Fe, respectively, in different types of supernovae. A full GCE approach is required as $^{53}$Mn, $^{53}$Mn, and $^{56}$Fe are produced by both CCSNe and SN Iae, while $^{60}$Fe originate exclusively from CCSNe. As supernova yields are uncertain, and the exact fraction of SN Ia that have a sub-Chandrasekhar mass progenitor is unknown, we tested the impact of different yields and different values of this fraction. 

We found that no combination of yields, SN Ia DTD, or value of $K$ lead to isolation times for $^{53}$Mn and $^{60}$Fe that have maximum likelihood less than $5$ Myr of each other. We consider also a scenario where SN Ia are assumed to contribute no $^{53}$Mn in the ESS, and find several solutions where the likelihood of the four distributions studied so far in detail with GCE ($^{53}$Mn, $^{60}$Fe and the $s$-process SLRs $^{107}$Pd and $^{182}$H) overlap significantly. The best cases, that is, where the peaks of $^{107}$Pd and $^{182}$Hf overlap with both the $^{53}$Mn and $^{60}$Fe distributions at a probability density of around 0.1, are generally found for $K=1.6$ and 2.3 for the LC18(setR) and N13 CCSN yields. 

Recently, \cite{Fang:2025} published a new value for the ESS $^{60}$Fe/$^{56}$Fe ratio of $7.71\pm0.47\times10^{-9}$. Compared to $10.1\pm2.7\times10^{-9}$ from \cite{Tang:2015} we adopted here, the new value is roughly 30\% lower, resulting in a shift in the distribution of +1 Myr, alleviating the $\sim5$ Myr gap between the maximum likelihoods of the $^{53}$Mn and $^{60}$Fe distributions. The new value is also five times more precise, significantly narrowing the KDE distributions, so that the probability density peak for $^{60}$Fe reaches 0.29 instead of 0.24. In any case, the general conclusions drawn above do not change significantly. 

In conclusion, although not the most likely, our results indicate that a self-consistent origin scenario for $^{53}$Mn, $^{60}$Fe, $^{107}$Pd, and $^{182}$Hf in the ESS is possible when considering a purely GCE origin. The likelihood of this scenario depends on the specific value of $K$ and the choice of CCSN yields, but not on the choice of DTD for the sub-$M_{\text{Ch}}$ population. The Revised GCE models with $t_{\text{max}}=4.3$ Gyr have the largest discrepancy between the $^{53}$Mn and $^{60}$Fe isolation times, while the GCE runs that assumed no $^{53}$Mn contribution in the ESS from SN Ia led to better agreement with the isolation time distributions of the other SLRs. The most likely value of the isolation time is in the range between 9 and 12 Myr. 
%except for $K=5.7$ where higher values of $f_{\text{sub}}$ are necessary to fit the $^{182}$Hf distribution. 

Our conclusions do not support the need for a local source of these SLRs, as their abundances can be already accounted for by the operation of the GCE. This is unless the isolation time was so long (e.g., $>$ 50 Myr for \iso{53}Mn and $>$100 Myr for \iso{182}Hf for a 10,000 decrease factor) that most of the GCE contribution would have been decayed before the formation of the first solids. Furthermore, the GCE scenario presents some likelihood of explaining the SLRs self-consistently, while a local CCSN scenario so far has not managed to obtain such a result \citep{vesconi18,Battino:2024}. Our results assume that $^{60}$Fe is produced only by CCSNe occurring at a typical rate consistent with observations; therefore, our assumptions are consistent with the proposed frequency of the production site necessary for $^{60}$Fe deep-sea detections \citep[e.g.,][]{Wallner:2021} as investigated by \citealt{Wehmeyer:2023} (see their Figure 2). Possible deep-sea excesses of $^{53}$Mn have also been found in the same time-window as $^{60}$Fe, which could indicate that the abundances of these SLRs were associated with the same supernova event \citep{Korschinek:2020}.

Future work involves considering the possible contribution to $^{60}$Fe from electron-capture supernovae \citep{jones:2019,wanajo13}, and Super-AGB stars \citep{lugaro:2012slrs, jones:2016}, which were not included here due to the large uncertainties related to their physics, and their still debated occurrence and rarity. Including them as an extra source of $^{60}$Fe may somewhat improve the likelihood of our solution and soften the problem related to the $^{60}$Fe/$^{53}$Mn ratio possibly related to the fraction of faint CCSNe, however, it would increase the predicted $^{60}$Fe/$^{26}$Al ratio, which is already typically higher than observed via gamma-ray flux \citep[e.g.][]{Siegert:2023}. Detailed GCE models of the origin of the $s$-only SLR \i\iso{205} Pb are also still required, as the results of \citet{Leckenby:2024} are based on the steady-state equation approximation. Another missing piece of the puzzle of the origin of SLRs in the ESS is related to the production of those SLR heavier than iron and proton rich, such as $^{92}$Nb and $^{146}$Sm. The origin of these nuclei is still debated \citep[e.g.,][]{Travaglio:2014,lugaro16,roberti23} and needs a future dedicated study. 

\begin{acknowledgements}
This work was supported by the European Union’s Horizon 2020 research and innovation programme (ChETEC-INFRA -- Project no. 101008324), the IReNA network by NSF AccelNet (Grant No. OISE-1927130), the European Union’s Horizon 2020 research and innovation programme (ERC-CoG-1026 Radiostar 724560), the Lend\"ulet Program LP2023-10 of the Hungarian Academy of Sciences, and the NKFI via K-project 138031 (Hungary). T.T. and M.P. acknowledge the support of the ERC Synergy Grant Programme (Geoastronomy, grant agreement number 101166936, Germany). M.L. was also supported by the NKFIH excellence grant TKP2021-NKTA-64. 

\end{acknowledgements}

\bibliography{main}{}
\bibliographystyle{aasjournal}

\begin{appendix}

\begin{figure*}[h!]
\centering
    \includegraphics[width=.32 \linewidth]{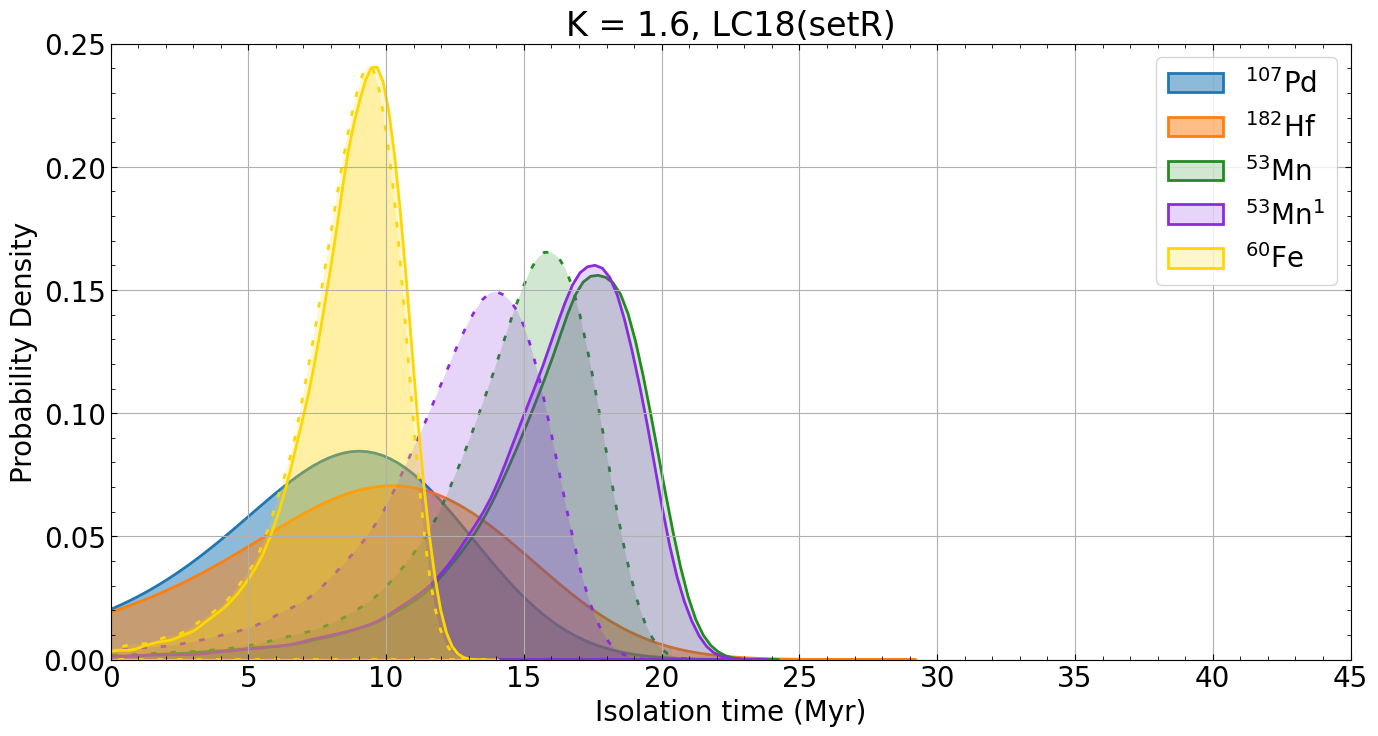}
    \includegraphics[width=.32 \linewidth]{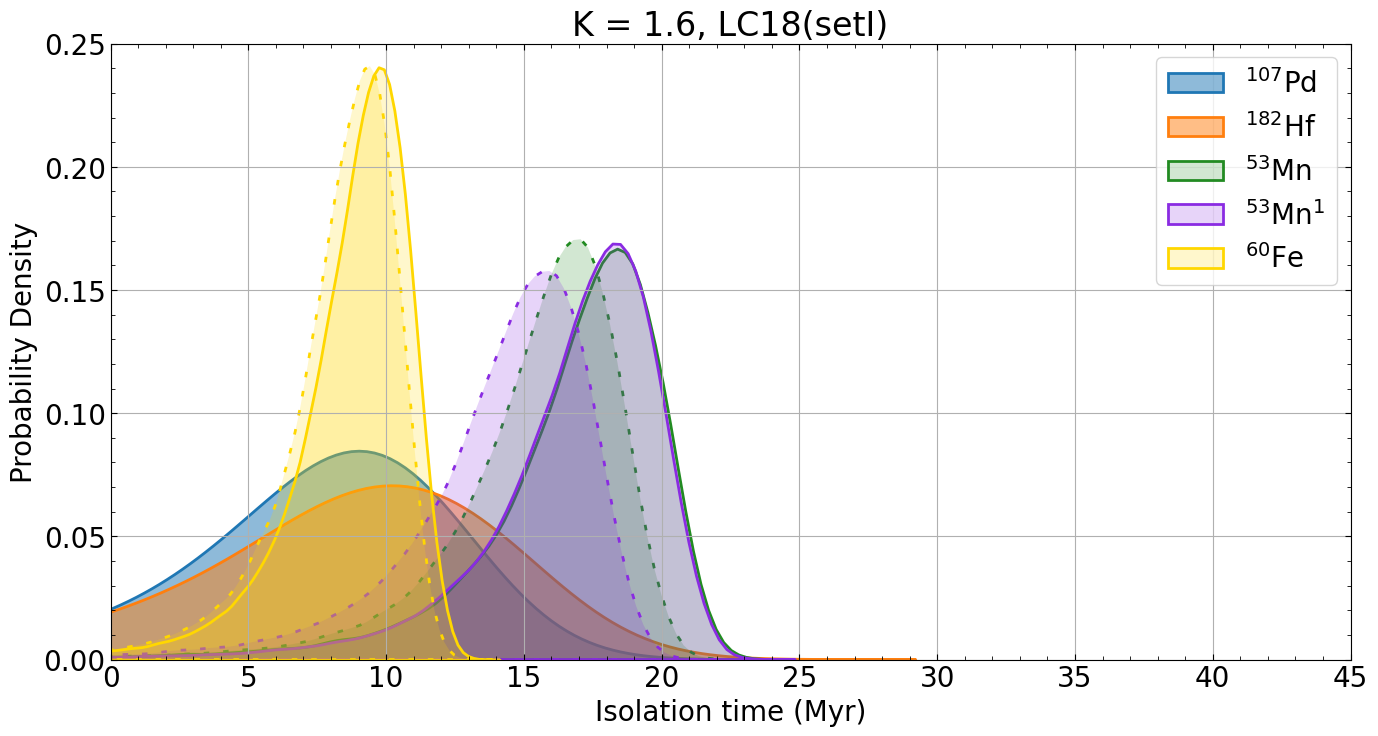}
    \includegraphics[width=.32 \linewidth]{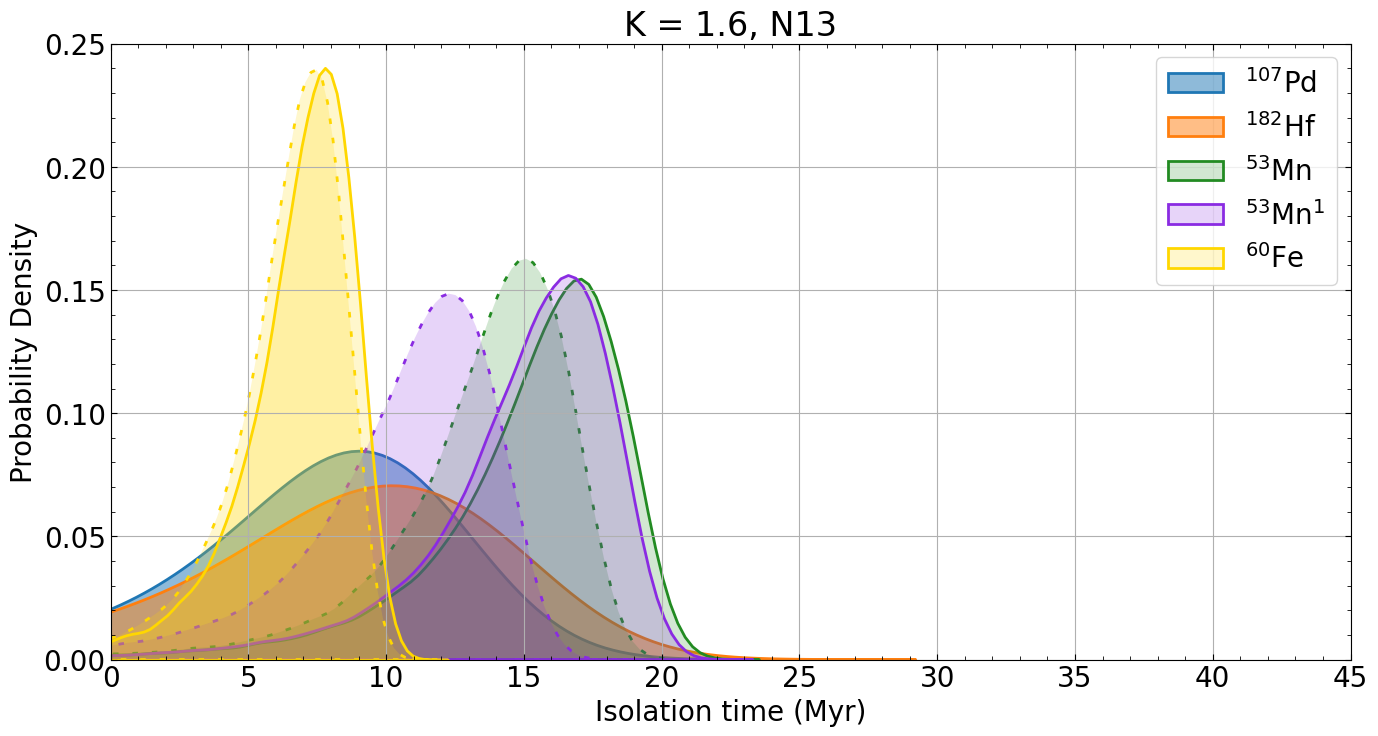}

    \includegraphics[width=.32 \linewidth]{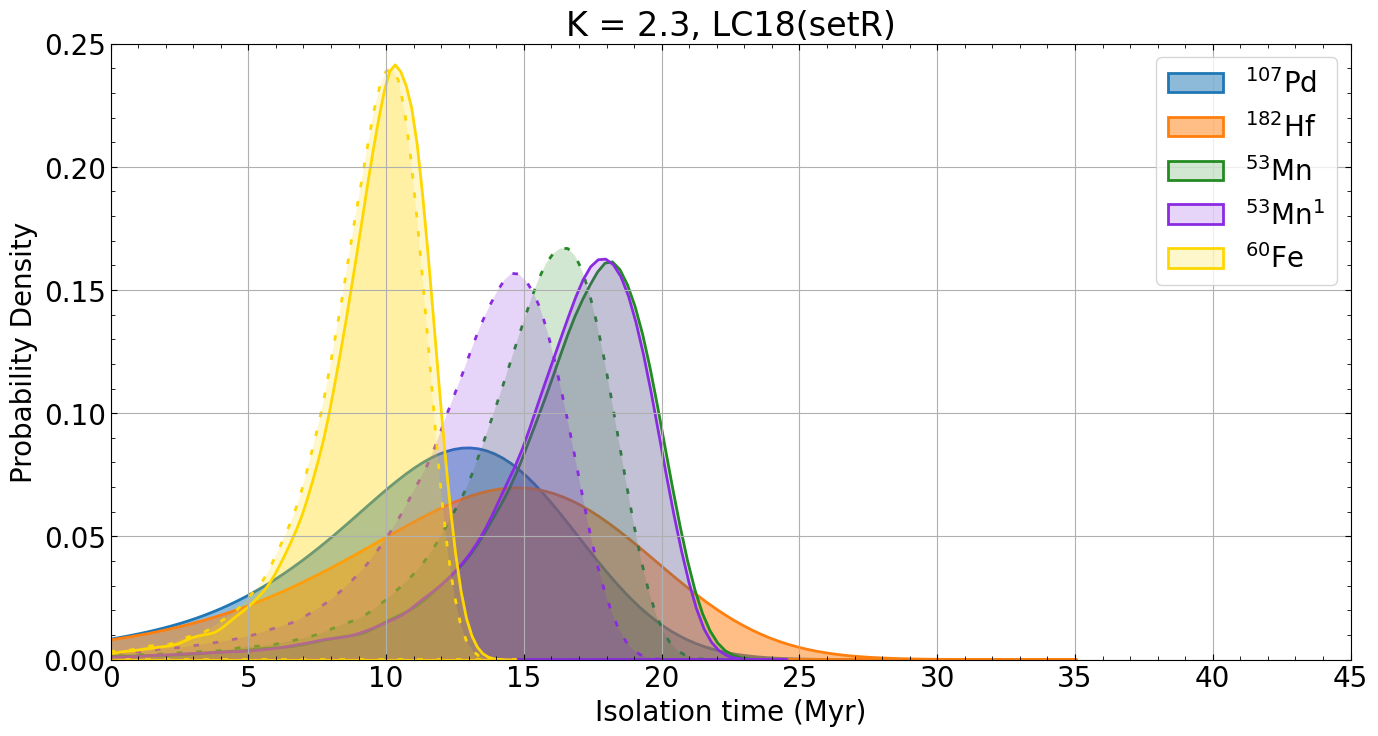}
    \includegraphics[width=.32 \linewidth]{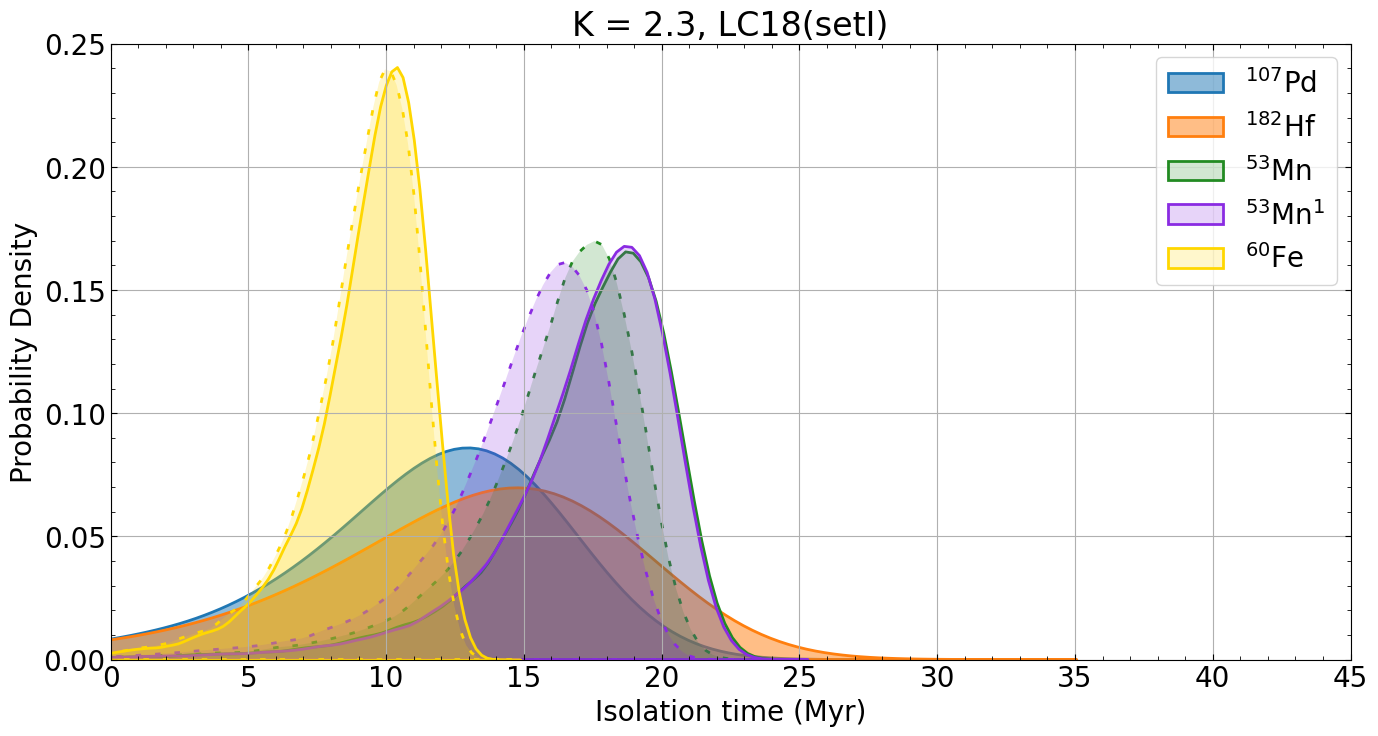}
    \includegraphics[width=.32 \linewidth]{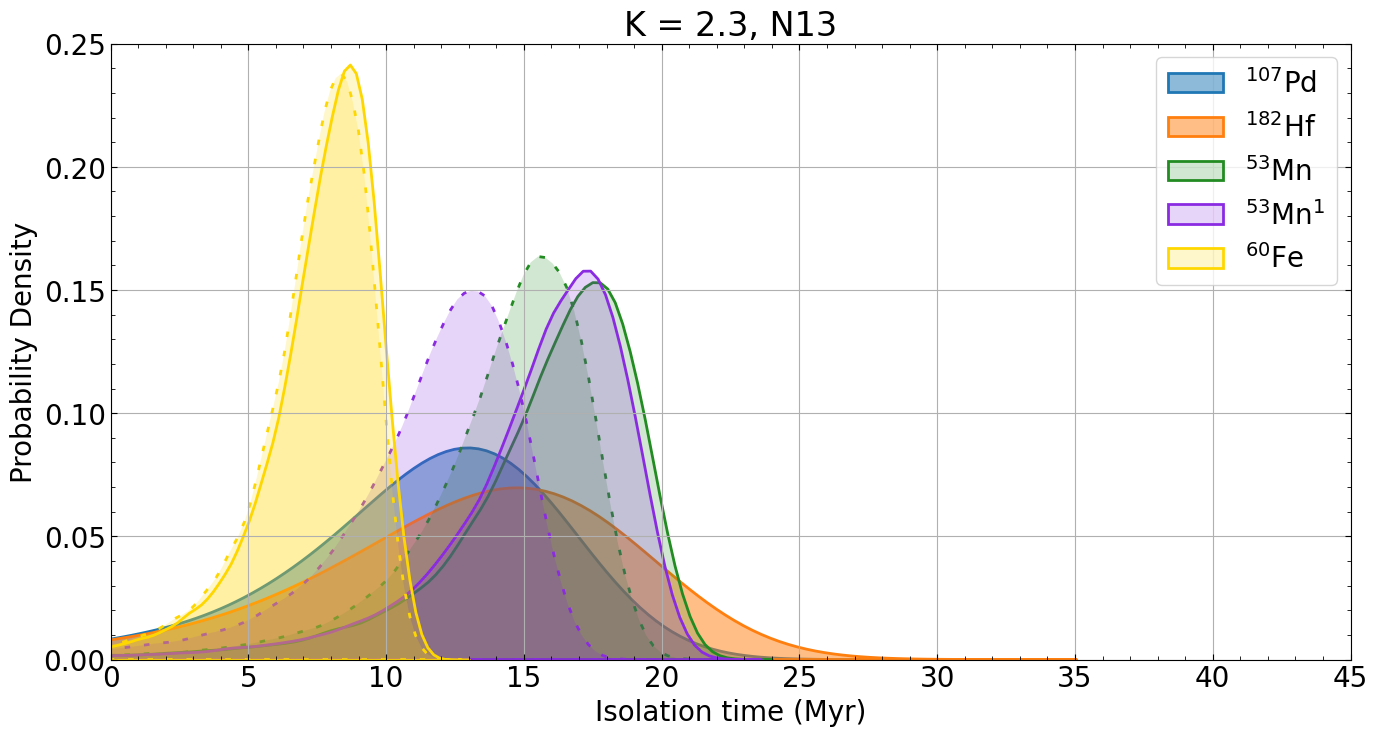}

    \includegraphics[width=.32 \linewidth]{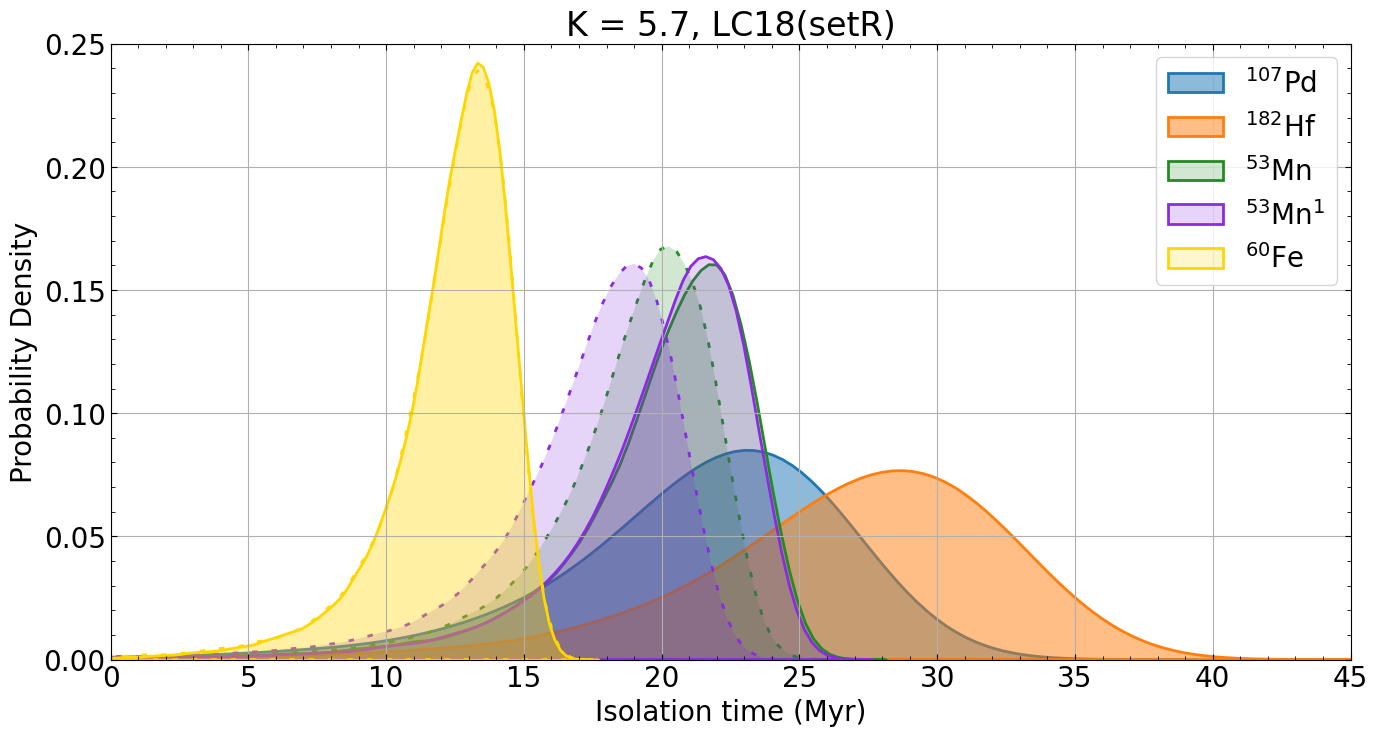}
    \includegraphics[width=.32 \linewidth]{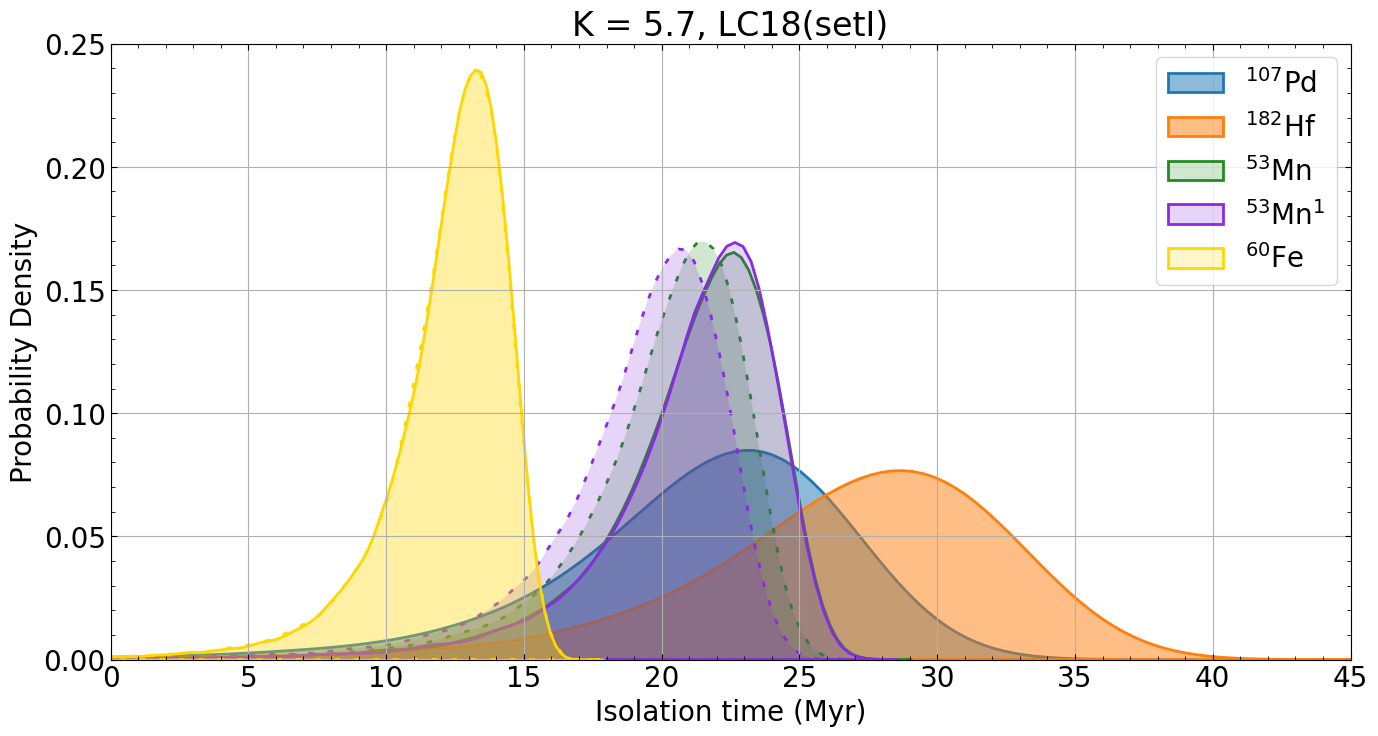}
    \includegraphics[width=.32 \linewidth]{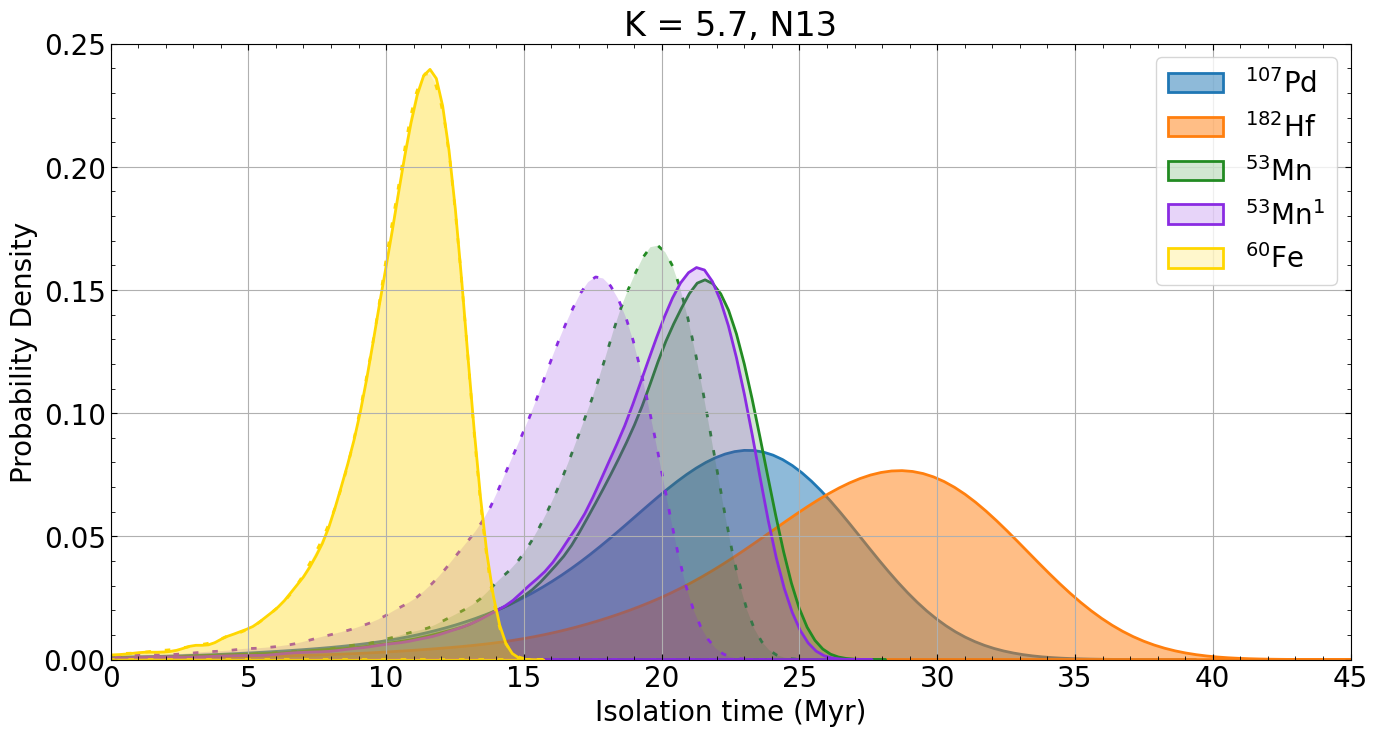}

    \includegraphics[width=.32 \linewidth]{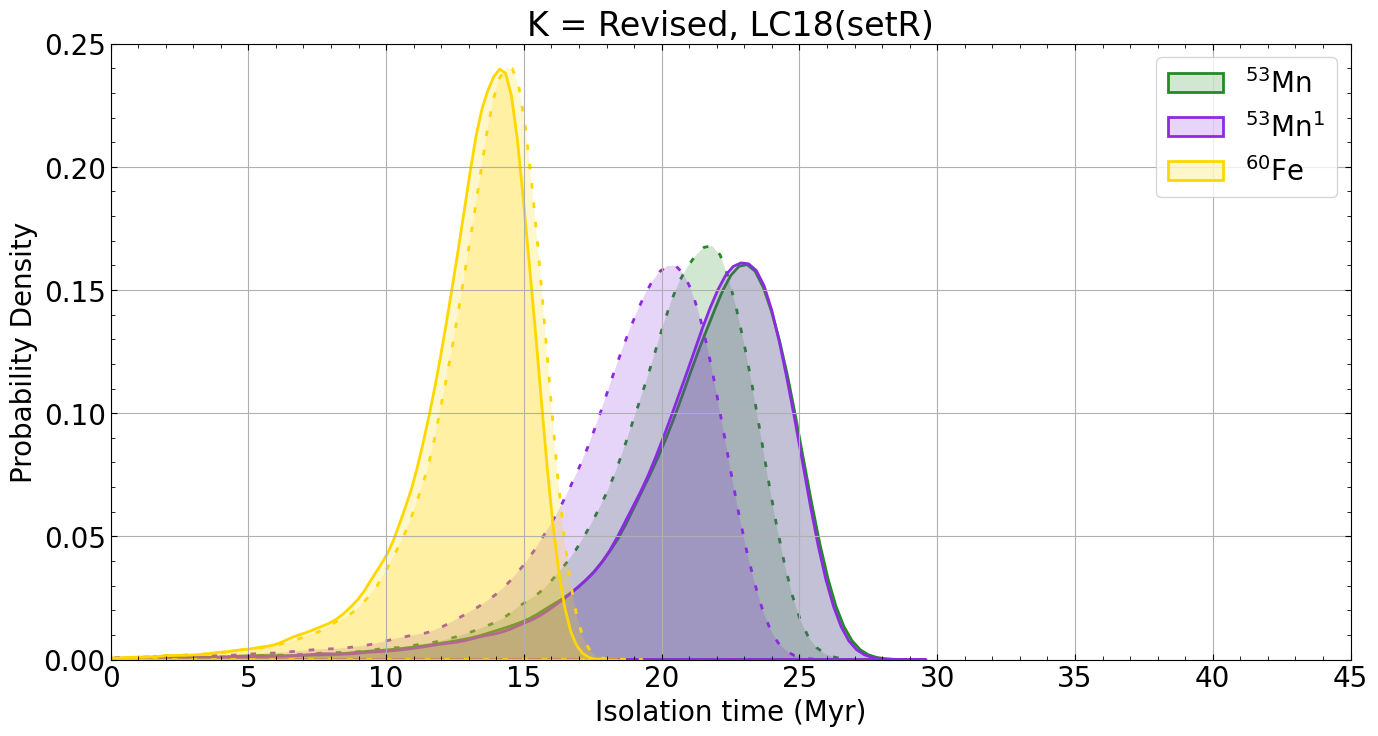}
    \includegraphics[width=.32 \linewidth]{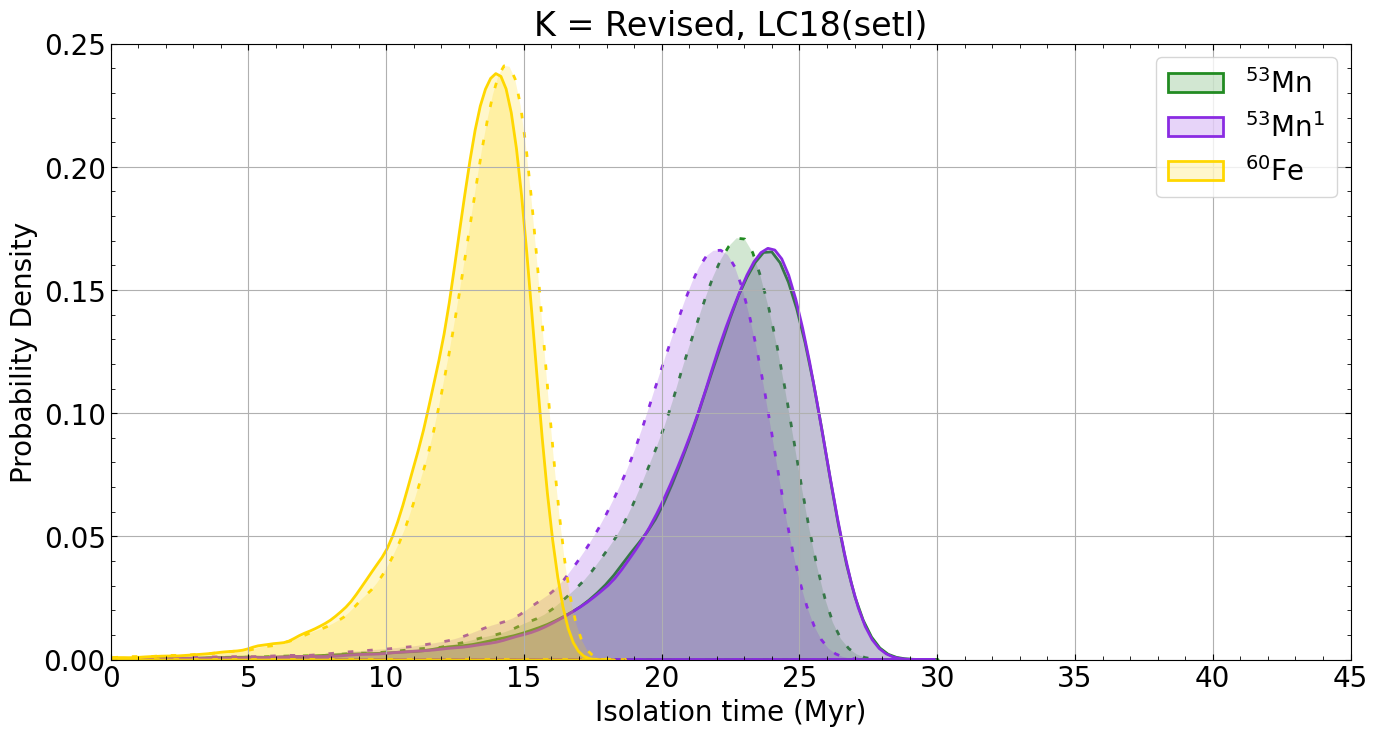}
    \includegraphics[width=.32 \linewidth]
    {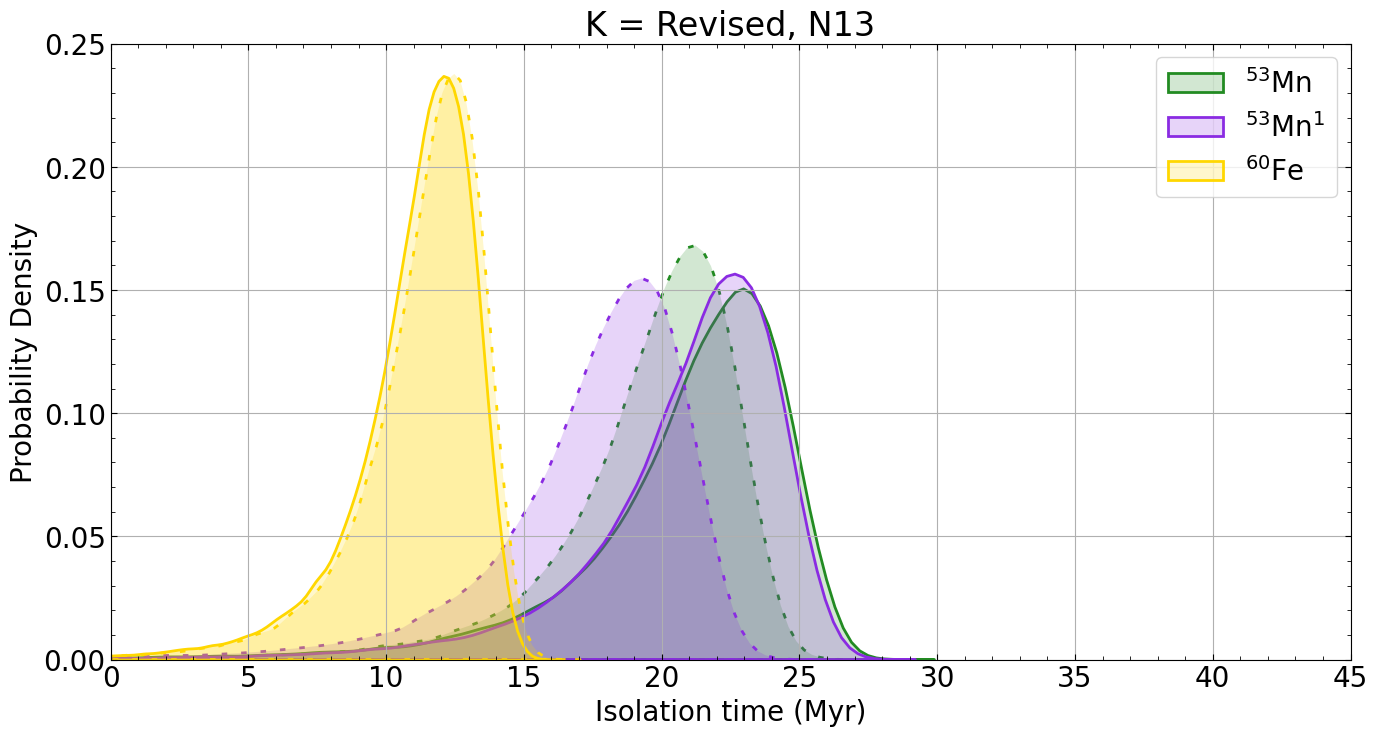}
    
\caption{Same as Figure \ref{fig:fig_norm_dist_tardy}, but for GCE models with a prompt sub-$M_{\text{Ch}}$ SN Ia DTD.}
\label{fig:fig_norm_dist_prompt}
\end{figure*}

\end{appendix}
\end{document}